\documentstyle{mn}

\title[IR mergers and IR QSOs with Galactic Wind. II. NGC\,5514]
   {IR mergers and IR QSOs with galactic winds. II.  \\
    NGC\,5514: two extra--nuclear starbursts with LINER\\
    properties and a supergiant bubble in the rupture phase\\
}

\author[L\'{\i}pari et al.]
    {S. L\'{\i}pari$^{1,2}$, E. Mediavilla$^{3}$, B. Garcia-Lorenzo$^{3,2}$,
    R.J. D\'{\i}az$^{1}$, J. Acosta-Pulido$^{3}$, 
   \newauthor
    M.P. Ag\"uero$^{1}$, Y. Taniguchi$^{4}$, H. Dottori$^{5}$, and R. Terlevich$^{6}$\\
$^{1}$ C\'ordoba Observatory and CONICET, Laprida 854, 5000 C\'ordoba, Argentina.\\
$^{2}$ Visiting astronomer at La Palma, CASLEO and CTIO Observatories.\\
$^{3}$ Instituto de Astrofisica de Canarias, 38205 La Laguna, Tenerife, Canary Island, Spain.\\
$^{4}$ Astronomical Institute, Tohoku University, Aoba, Senday 980-8578, Japan.\\
$^{5}$ Instituto de Fisica, Univ. Fed. Rio Grande do Sul, CP 15051, Porto Alegre, Brazil. \\
$^{6}$ Institute of Astronomy, Madingley Road, Cambridge CB3 0HA.
}

\date{Received     ;
      in original form }

\pagerange{\pageref{firstpage}--\pageref{lastpage}}
\pubyear{2002}

\begin{document}

\maketitle

\label{firstpage}

\begin{abstract}

A study of the morphology, kinematics and ionization structure of the IR merger
NGC\,5514, is here presented. This study is based mainly on INTEGRAL
two-dimensional (2D) spectroscopy (obtained at the 4.2 m William Herschel
Telescope, WHT), plus optical and near IR images.
Clear evidence of two extra-nuclear starbursts  with young outflows (OFs)
and LINER activity are reported.
One of these OFs has generated a supergiant bubble and the other
is associated with an extended complex of H {\sc ii} region.

In the galactic bubble it was found that:
(i) the [S {\sc ii}], H$\alpha$, [N {\sc ii}], [O {\sc i}] and [O {\sc iii}]
emission line maps show a bubble with a distorted ellipsoidal shape, with
major and minor axes of  $\sim$6.5 kpc (13\farcs6;
at PA $=$  120$^{\circ}\pm10^{\circ}$) and $\sim$4.5 kpc (9\farcs6);
(ii) these maps depict four main knots, a very strong one and 
three others more compact and located at the border;
(iii)\,the centre of the bubble is located at $\sim$4.1 kpc (8\farcs5) to the
west from the main nucleus;
(iv) the WHT spectra show---in this area---two strong components, blue and
red emission line systems, probably associated with emission from the
near and far side of the external shell, for which the
mean OF velocities were measured of
$V_{OF blue}=(-320\pm20)$\,km\,s$^{-1}$ and
$V_{OF red}=(+265\pm25)$\,km\,s$^{-1}$;
(v) these two components depict LINER properties, probably associated with
large scale OF + shocks;
(vi)\,at the east border, the kinematics of the ionized gas and the [S {\sc ii}]
emission line maps show an extended ejection of 4 kpc aligned
with the PA of the major axis;
(vii)\,another three ejections were found, two of them perpendicular
to the extended one.
Each ejection starts in one of the knots.
These results suggest that the bubble is in the rupture phase.

For the complex of giant H {\sc ii} regions it was detected that:
(i) the H$\alpha$, [N {\sc ii}] and
[S {\sc ii}] emission line maps show
a compact area of strong line emission line (peaking at
 $\sim$810 pc $\sim$1\farcs7, to the east of the second nucleus) and faint
extended emission with an elongated shape, and major and minor axes of
$\sim$5.1 kpc (10\farcs8; at PA $\sim$20$^{\circ}$)
and $\sim$2.9 kpc (6\farcs0);
(ii) inside this complex, the spectra show H {\sc ii} region and transition
LINER/H {\sc ii} characteristics; (iii)
at the border of this extended H {\sc ii} area the spectra have
outflow components and LINER properties.

INTEGRAL 2D [N {\sc ii}], H$\alpha$, [S {\sc ii}] and [O {\sc iii}]
velocity fields (VFs)  are presented.
These VF maps show results consistent with an expansion
of the bubble, plus four ejections of ionized gas.
The U, B, V, I, J, H, and K$_S$ images show a pre$-$merger morphology, from
which faint filaments of emission emerge, centred on the bubble.
The ionization structure and the physical conditions were analysed
using the following 2D emission line ratio and width maps:
[S {\sc ii}]/H$\alpha$, [N {\sc ii}]/H$\alpha$,
[O {\sc i}]/H$\alpha$, [O\,{\sc iii}]/H$\beta$ and FWHM--[N {\sc ii}].
In the region of the bubble, 100 per cent of the
[N\,{\sc ii}]/H$\alpha$ and [S {\sc ii}]/H$\alpha$ ratio show very high values
($>$ 0.8) consistent with LINER processes associated with fast
velocity shocks.
These new results support the previous proposition that extreme nuclear
and `extra-nuclear' starbursts with galactic winds + shocks play an
important role in the evolution of IR mergers/QSOs.

\end{abstract}


\section{Introduction}\label{introduction}

A current key issue in astrophysics is to explore the evolution of the
star formation process, especially at high redshift, when the galaxies/QSOs
formed, and where it is expected that the star formation rate is very high.
An important step, in order to improve our understanding of this issue, it
is to study extreme star formation and associated galactic wind processes
in nearby galaxies, since we can obtain more detailed and unambiguous data
(L\'{\i}pari et al. 2004a, 2003; Taniguchi et al. 2004, in preparation).
Mainly for this purpose we have conducted a study of infrared (IR) mergers
and IR QSOs with galactic winds (L\'{\i}pari et al. 2004a, 2003, 2000;
L\'{\i}pari, Colina \& Macchetto 1994; L\'{\i}pari 1994); these IR systems are
excellent laboratory at low redshift for the study of extreme star formation
and galactic wind processes.

\subsection{IR mergers and IR QSOs}\label{intrmq}

One of the most exciting population of objects identified
over the last two decades is the group of IR bright galaxies, which emit
most of their energy in the IR (L$_{IR[8-1000 \mu m]}$/L$_{B} \sim$ 5--300;
Soifer et al. 1984; Rieke et al. 1980). In particular,
luminous  and ultraluminous IR galaxies
(LIRGs: L$_{IR} \geq 10^{11} L_{\odot}$ and ULIRGs: L$_{IR} \geq
10^{12} L_{\odot}$, respectively) are
dusty, strong IR emitters where frequently a strong enhancement of star
formation is taking place (for references see  L\'{\i}pari et al. 2003, 2004a).
In addition, imaging surveys of LIRGs and ULIRGs
show that a very high proportion ( $\sim$70--95$\%$) are mergers or interacting
systems (Joseph \& Wright 1985; Rieke et al. 1985; Sanders et al. 1988a;
Melnick \& Mirabel 1990; Clements et al. 1996).
These IR galaxies also contain large amounts of centrally concentrated
molecular gas (Sanders, Scoville \& Soifer 1991;
Scoville et al. 1991).
There is observational evidence and theoretical works suggesting that in
IR mergers/QSOs tidal torque and loss of angular momentum drive large amount
of interstellar gas into the central regions, leading to {\it extreme starburst
processes} and probably fuelling a supermassive black hole
(see  Joseph \& Wright 1985; Rieke et al. 1985; Heckman, Armus \& Miley 1987,
1990; Sanders et al. 1988a,b; Scoville \& Soifer 1991; Barnes \& Hernquist
1992, 1996; Mihos \& Hernquist 1994a,b, 1996; Taniguchi \& Wada 1996;
Canalizo \& Stockton 2001).

At the highest IR luminosities, the presence of AGN in LIRGs becomes
important, and LIRGs probably represent an important stage in the
formation of QSOs and elliptical galaxies. In particular,
several possible links among mergers, starbursts, ellipticals and QSOs have
been proposed (e.g., Joseph \& Wright 1985; Sanders et al. 1988a; L\'{\i}pari 1994).
The discovery and study of IR QSOs (see L\'{\i}pari et al.
2003, 2004a; Zheng et al. 2002) raises several interesting
questions, in particular whether they are a special class of QSOs.
We found, or confirm, that a high percentage of IR QSOs show extreme galactic
winds with giant galactic arcs, merger features, BAL systems, and extreme
Fe {\sc ii} emission, and are radio quiet (L\'{\i}pari et al. 2003, 2004a;
L\'{\i}pari, Terlevich \& Macchetto 1993; L\'{\i}pari, Macchetto \& Golombek
 1991a).
We suggested that these objects could be {\it young IR$-$active
galaxies at the end phase of a strong starburst: i.e., composite and
transition QSOs}.

\subsection{Galactic winds and galactic bubbles}\label{intrwb}

Galactic winds (GWs) and outflows (OF) have been detected mainly in starburst and Seyfert
galaxies (see Heckman et al. 2000; Veilleux, Kim \& Sanders 2002a).
IR mergers/QSOs often show strong and extreme starbursts, with very
powerful galactic winds (L\'{\i}pari et al. 2003, 2004a).
In addition, galactic shells, arcs and bubbles generated by multiple SN
explosion and massive star winds have been found in the Milky Way, the Magellanic
Clouds, M 31, M 33 and other nearby galaxies.
However, there are few examples of GWs associated with extreme starburst
detected in their early phases, i.e. in the `supergiant' galactic bubble stage.
Furthermore, mainly broken supergiant bubble/shells
in the blowout or post$-$blowout (free wind) phases have been detected: e.g.
from the nuclei of NGC 3079, Arp 220, Mrk 231, IRAS 19254--7245, NGC 2623,
NGC 2782, etc. (Hummel, van Gorkon \& Kotanyi 1983; Ford et al. 1986;
Duric \& Seaquist 1988; Heckman et al. 1987, 1990; Hodge \& Kennicutt 1983;
Lipari et al. 1994, 2003, 2004a; Jogee, Kenney \& Smith 1998, and others),
from the extra-nuclear regions of  NGC 6946, NGC 1620, etc. (Hodge 1967;
Efremov et al. 2002; Vader \& Chaboyer 1995, etc.),
and also from both the nucleus and
extra-nuclear areas of M82 (Lynds \& Sandage 1963; Axon \& Taylor 1978;
Bland \& Tully 1988; Wills et al. 1999; Garcia-Burillo et al. 2001).
Thus, young extreme GWs with supergiant galactic bubbles (even in the
blowout phase) are difficult to detect because of their
short timescale of $\leq$ 10$^7$ yr (Suchkov et al. 1994).

Our understanding of the main phases of galactic winds associated with
starbursts was improved significantly by the use of theoretical and
numerical models (see Strickland \& Stevens 2000; Suchkov et
al. 1994; Mac Low, McCray \& Norman 1989; Tomisaka \& Ikeuchi 1988).
In general, good agreement has been found between these  models and the
observations (L\'{\i}pari et al. 2004a).
Theory suggests four main stages for GWs associated with starbursts (following
the review presented by Lehnert \& Heckman 1995, 1996; Heckman et al. 1990):

\begin{itemize}

\item
{\it Phase I}: A GW results when the kinetic energy of the ejecta supplied by
multiple  supernovae and winds from massive stars is high enough to excavate
a cavity in the centre of a starburst.
At this point the kinetic energy is converted into thermal
energy. This conversion means that the collision of
intersecting SN/stellar winds transforms the kinetic energy of the ejecta into
thermal energy via {\it shocks}. In this cavity the host fluid (10$^8$ K)
has a sound speed much greater than that of the local escape velocity and a pressure
much higher than the ISM and thus will expand as a {\it `galactic bubble'}.

\item
{\it Phase II}: As the bubble expand and sweeps up the ambient gas, it will
enter the `radiative phase' (Castor, McCray \& Weaver 1975). The
bubble will  then collapse--due to radiative cooling--into a
{\it `thin shell'}.

At this phase there are different radial components in the GW:
(1) the inner starburst region where the energy is injected and thermalized;
(2) a region of supersonic wind;
(3) a hot region (T $\sim$10$^8$ K) where the wind gas has been decelerated
and heated by internal shocks; and
(4) the thin shell, which is the source of optical emission lines
(with velocities of several hundred km s$^{-1}$).

\item
{\it Phase III}: After the shell has formed its evolution is strongly
dependent on the input physics. If the cooling rate in the interior is high,
then the expanding bubble can stop expanding (Tomisaka \& Ikeuchi 1988).

\item
{\it Phase IV}: If other probable dynamical and thermal conditions are
considered (e.g. Suchkov et al. 1994; MacLow et al. 1989), the shell can
{\it `break up'}. After this break up the host interior become a freely
expanding wind, and the bubble then {\it `blows out'}.

In the blow out phase the optical emission comes from obstacles,
such as clouds and  shell fragments, which are immersed and shock--heated
by the
OF. In addition, in this phase of free expansion, the velocity, pressure,
temperature and density of the wind are a function of the radius.

\end{itemize}

On the other hand, for galactic winds associated with AGN the situation
is less clear, and very different models are proposed in order to explain
the observed data. In these models the OF could be generated by
jets driven thermal winds, accretion discs winds, X-ray heated torus
winds, etc. (see Veilleux et al. 2002a;
Morganti et al. 2003).

In addition, it is important to remark that in recent years the development
of new technology has allowed  the study of GWs and outflow  processes
for almost all the main components of the ISM, including the ionized warm gas,
the hot gas (10$^7$ K), cold neutral H {\sc I}, molecular gas and dust
(e.g., Heckman et al. 1996, 2000; Lipari et al. 2000, 2003, 2004a; Wills et
al. 1999; Oosterloo et al. 2000; Seaquist \& Clark 2001; Alton, Davies \&
Bianchi 1999; Morganti et al. 2003).

\section{THE PROGRAMME AND OBSERVATIONS} \label{pobservations}

\subsection{The programme} \label{intrp}

One of the most important astrophysical issues in modern astronomy is
to explore the formation and early evolution of galaxies at high redshift.
To improve our understanding of this issue, it is necessary to
study: (i) objects with strong star formation and galactic
wind processes at low redshift (Lipari et al. 1994, 2000, 2003, 2004a);
(ii) well-defined samples of forming + GW galaxies at high redshift
and then investigate their nature (e.g. Taniguchi \& Sioya 2000;
Taniguchi et al. 2003; Ajiki et al. 2002, 2003).
With these aims in mind we began an international project to study nearby star forming
+ GW galaxies and distant Ly$\alpha$ emitters (see Lipari
et al. 2004a,b; Taniguchi et al. 2003; Ajiki et al. 2002, 2003).

The first step in this project is to understand the star formation process
in nearby galaxies because we can obtain more detailed and unambiguous
information. Thus, our groups started a  study of nearby IR mergers/QSOs,
which are an excellent laboratory--at low redshift--for the analysis of
extreme star formation and GW processes (Lipari et al. 2004a). In
the present paper we show  new results from this study of the
morphology, kinematics and ionization structure of `nearby' IR mergers/QSOs
with galactic winds.
This study is based mainly on two-dimensional spectroscopy, obtained at
the European Northern Observatory (ENO, La Palma--Spain), the European Southern
Observatory (ESO, Chile), the Complejo Astronomico El Leoncito (CASLEO,
Argentina), and Bosque Alegre (BALEGRE, Argentina) observatories, with the
4.2 m, 3.6 m, 2.15 m and 1.5 m telescopes, respectively.
The characteristics and goals of the programme have been
described in detail by L\'{\i}pari et al. (2004a). The first $\sim$10 objects,
observed with 2D spectroscopy, include nearby IR systems selected from our
previous surveys and from the literature.
Our original database of IR mergers/QSOs with outflow  and candidates
contains a total of 43 objects (see Table 1 of L\'{\i}pari et al. 2004a).

The main goal of this programme (at low redshift) is to analyse in
detail the properties of the different stages of extreme starbursts, galactic
winds, mergers, QSOs, and elliptical galaxies (and their interrelation).
NGC\,5514 is one of the ten nearby IR systems of our first 2D spectroscopic
study (and it was also included in the database of 43 IR mergers/QSOs
with outflow; L\'{\i}pari et al. 2004a).
This object was originally selected from the {\it `Survey of Warm IRAS
Galaxies Candidates'} (L\'{\i}pari, Bonatto \& Pastoriza 1991b; L\'{\i}pari,
Macchetto \& Golombeck 1991a; L\'{\i}pari \& Macchetto 1992a,b). This
Survey is based on observations of AGN/QSOs and starburst candidates, from
the {\it `Catalogue of Warm IRAS Sources'} (de Grijp et al. 1987).
In this pre-merger system (NGC\,5514) we found two strong `extra-nuclear'
starbursts, with two associated early outflows. This object therefore
gives important clues  about the early phases of the processes that we
enumerated as the main goals of this programme at low redshift.
Furthermore,  NGC\,5514 shows two new examples of young GWs with LINER
properties, and probably the first GW observed in the `pre-blowout phase'
of a supergiant extra-nuclear galactic bubble.
We note that this OF structure shows properties similar to those
observed in the supergiant galactic bubble of NGC 3079.
In NGC 5515 however, this OF bubble is located far from the nuclear region
(and hence far from a possible AGN) and is just in the rupture process.

NGC\,5514 (UGC 9102, IRAS 14111+0753, de Grijp et al. Catalogue N--342) is a
nearby IR galaxy at z$_{sys}$ = 0.024527,
which is the result of a merger between probably two disc galaxies of
unequal mass (M$_1$/M$_2$ $\sim$ 2; Fried \& Lutz 1988, hereafter FL88).
This system shows two compact nuclei, and a bright tidal tail (see Section 3.1).
The total IR luminosity of NGC\,5514 is
L$_{IR[8-1000 \mu m]} \sim 0.5 \times 10^{11} L_{\odot}$.
Throughout this paper, a Hubble constant of H$_{0}$ = 75 km~s$^{-1}$
Mpc$^{-1}$ will be assumed.
For NGC\,5514 a distance of $\sim$98.1 Mpc (cz$_{sys}$ = 7358
$\pm$ 25 km~s$^{-1}$; Section 3.5) was adopted, and thus the angular scale is
1$'' \approx$ 476 pc.

\subsection{WHT + INTEGRAL 2D spectroscopy} \label{o2ds}

In general,
the observations were obtained at ENO, CASLEO and CTIO
with the 4.2, 2.15 and 1.0 m telescopes, respectively.

Two-Dimensional (2D) optical spectroscopy of NGC\,5514 was obtained
during two photometric nights in April 2001 at the 4.2 m William Herschel Telescope (WHT)
at the Roque de los Muchachos Observatory on the island of La
Palma, Spain (Table~\ref{observations}).
The WHT was used with the INTEGRAL
fibre system (Arribas et al. 1998) and the WYFFOS spectrograph
(Bingham et al. 1994). The seeing was typically  1\farcs0.

INTEGRAL links the f/11 Nasmyth focus of the WHT with the slit of WYFFOS
via three optical fibre bundles. A detailed technical description of INTEGRAL
is provided by Arribas et al. (1998); here we recall only its main
characteristics. The three bundles have different spatial
configurations on the focal plane and can be interchanged depending on the
scientific programme. At the focal plane the fibres
of each bundle are arranged in two groups, one forming a rectangle and the
other an outer ring (for collecting background light, in the case of
small objects).

The data analysed in this paper were obtained with the standard bundle 2 of
219 fibres, each 0\farcs9 in diameter projected on the sky.
The central rectangle is formed by 189 fibres, covering a field of view of
16\farcs4 $\times$ 12\farcs3; and the other 30 fibres form a ring 90$''$
in diameter.

The WYFOS spectrograph was equipped with two gratings of 1200 line mm$^{-1}$,
covering the $\lambda\lambda$6000--7400 and $\lambda\lambda$4500--5900 \AA\,
ranges. A TEK CCD array of
1124 $\times$ 1124 pixels of 24 $\mu$m side was used, giving a linear
dispersion of about $\sim$1.4 \AA\ pixel$^{-1}$ ($\sim$2.8\,\AA\, effective
 resolution, $\sim$ 100 km s$^{-1}$).
Using the red grating, three different positions of the central region of this
merger were observed (called Positions 1, 2 and 3; see Section 3.2 and Table 1)
in order to obtain large spatial coverage. These three observed regions were
overlapped, forming a mosaic and covering a total area of
approximately 30$'' \times$ 20$''$. We note that the main nucleus of the merger was
observed in all the positions, in order to perform a simple
overlapping process.
Positions 1 and 3 were also observed using the blue grating.

It is important to note that for this mosaic of 2D observations, our aim
was to cover the three main central regions of the merger: two extra-nuclear
areas with strong optical emission lines and the nuclear area (for
details see Sections 3.2 and 3.3). Thus, in the present work (using
INTEGRAL 2D spectroscopy), our aim was not to cover the entire main
body  of NGC 5514.

\subsection{CASLEO long--slit + aperture spectroscopy and broad band images}
\label{ols}

Spectrophotometric observations and images of NGC\,5514 were taken at CASLEO
(San Juan, Argentina) with the 2.15 m Ritchey-Chr\'etien telescope.
Optical long-slit + aperture spectroscopy and broad-band CCD imaging
observations were obtained during seven photometric nights in June 1989,
July 1993, March 1997 and May 2000 (see Table~\ref{observations}).
Long-slit  spectroscopic observations with medium and high
resolution were obtained with the University of Columbia spectrograph (UCS;
e.g. L\'{\i}pari et al. 1997).
The medium resolution spectra were made using a 600 line mm$^{-1}$ grating,
a slit width of 2\farcs5,
which gives an effective resolution of $\sim$6 \AA\ ($\sim$290 km s$^{-1}$)
and a dispersion of  120 \AA\ mm$^{-1}$, covering the wavelength range
$\lambda\lambda$4000--7500 \AA.
The high resolution spectra were made using a 1200 line mm$^{-1}$ grating,
a slit width of 2\farcs0,
which gives an effective resolution of $\sim$1.5 \AA\ ($\sim$50 km s$^{-1}$)
covering the wavelength ranges $\lambda\lambda$4700--5800 and
6100--7200 \AA.
Aperture spectroscopic data were obtained with the `Z-machine'
(e.g. L\'{\i}pari et al. 1991a,b).
These aperture spectra were made using a 600 line mm$^{-1}$ grating,
giving a dispersion of 130 \AA\ mm$^{-1}$ and an effective resolution
of $\sim$8 \AA\ ($\sim$300 km s$^{-1}$)
covering the wavelength range $\lambda\lambda$4700--7200.
U, B, V, and I images were obtained. A TEK 1K chip with a
scale of 1\farcs01 pixel$^{-1}$ was used.
The seeing was in the range 1\farcs5--2\farcs5 (FWHM).

\subsection{CTIO long-slit spectroscopy}
\label{olsctio}

Spectrophotometric observations were taken at the Cerro Tololo
Inter-American Observatory (CTIO, Chile) with the two-dimensional
photon-counting detector (2D FRUTTI) attached to the Cassegrain
focus of the 1.0 m telescope.
The observations were obtained during one photometric
night in March 1990 (see Table~\ref{observations}).
The medium resolution long-slit spectra, were taken with a slit width of
1\farcs5. The wavelength range was $\lambda\lambda$3600--7000 \AA.
From comparison lamp and sky lines an effective resolution
of $\sim$6--7 \AA\ ($\sim$300 km s$^{-1}$) was estimated.
The seeing was in the range 1\farcs3--1\farcs8 (FWHM).

\subsection{NED archive data} \label{ned}

From NASA Extragalactic Database (NED, California Institute of Technology)
we have obtained a copy of Palomar 48 inch Schmidt telescope image of
NGC\,5514. This deep photographic image was observed using a plate 103a-O
(broad band $\sim$B).

In addition, at NED and from the `Two Micron All Sky Survey' (2MASS;
Jarret et al. 2000) near IR images {\it J (1.2 $\mu$m), H (1.6 $\mu$m) and
K$_S$ (2.2 $\mu$m)} were obtained.
These images were observed using the 2MASS dedicated 1.3 m telescope, giving
 an angular resolution of $\sim$2$''$.

The California Institute of Technology
gave us authorization to use and show these NED archive public domain data.

\subsection{Reduction} \label{reduc}

The {\sc IRAF}\footnote{{\sc IRAF} is a reduction and analysis software
facility developed by NOAO.} software packages were used to reduce and analyse
the data (obtained at the ORM, CASLEO and CTIO).
The long-slit spectra and broad band images --obtained at CASLEO and
CTIO-- were reduced in the standard way. Bias and dark subtraction and
flat fielding were performed. Wavelength calibration of the
long-slit spectra was done by fitting two-dimensional polynomials to the
positions of lines in the arc frame. The  spectra were corrected
for atmospheric extinction, galactic reddening and redshift. The images
and the long-slit spectra were flux-calibrated using observations of standard
stars from the samples of Oke (1990), Landolt (1992) and Stone \& Baldwin
(1983).

The reduction of the 2D spectroscopic observations consists of two mains
steps: (i) reduction of the spectra for each of the 219 fibres and (ii)
the generation of 2D maps by interpolating the selected parameter (e.g.
emission line intensity, continuum intensity, radial velocity, etc.) from the
grid values defined by the fibre bundle.
Step (i) was basically done in the same way as for long-slit spectroscopy,
including bias subtraction, aperture definition and trace, stray light
subtraction, the extraction of the spectra, wavelength calibration, throughput
correction and cosmic-ray rejection.
We obtained typical wavelength calibration errors of 0.1 \AA, which give
velocity uncertainties of 5 km s$^{-1}$.
For step (II) the {\sc INTEGRAL}\footnote{{\sc INTEGRAL} is an imaging and spectroscopic
analysis software facility developed by the Instituto de Astrofisica de
Canarias (IAC).} software package was used
with 2D interpolation routines.
In particular, ASCII files with the positions of the fibres and the
corresponding spectral features, were transformed into regularly spaced
standard FITS files.

To generate two-dimensional maps of any spectral feature (intensity,
velocity, width, etc.) the INT-MAP script of the INTEGRAL
reduction package and the IDA tool (Garc\'{\i}a-Lorenzo, Acosta-Pulido, \&
Megias-Fernandez 2002) was used.
We have found that the IDA package gets better
results recovering 2D maps from low signal-to-noise data. The IDA
interpolation is performed using the IDL standard routine TRIGRID, which
uses a method of bivariate interpolation and smooth surface fitting for
irregularly distributed data points (ACM Transactions on Mathematical
Software, 4, 148-159).
Maps generated in this way are presented in the following sections.

The emission line components were measured and
decomposed using Gaussian profiles by means of a non-linear
least-squares algorithm described in Bevington (1969). In
particular, we used the software {\sc SPECFIT}\footnote{{\sc SPECFIT} was
developed and is kindly provided by Gerard A. Kriss.}, and SPLOT from the
{\sc STSDAS}\footnote{{\sc STSDAS} is the reduction and analysis software
facility developed by the Space Telescope Science Institute.}, and IRAF
packages, respectively.
An example of SPECFIT deblending, using three components for each emission
line in IRAS\,01003$-$2238, was shown in figure 2 of L\'{\i}pari et al.\ (2003). We note that in each WHT spectrum the presence of OF components and
multiple emission line systems were confirmed by detecting these systems
 in at least two or three different
emission lines ([N {\sc ii}]$\lambda$6583, H$\alpha$, [N {\sc ii}]\
$\lambda$6548, [S {\sc ii}]$\lambda\lambda$6717/31,
[O {\sc i}]\ $\lambda$6300, [O {\sc iii}]\ $\lambda$5007, and H$\beta$).
For the study of the kinematics,  the
{\sc ADHOC}\footnote{{\sc ADHOC} is a 2D/3D kinematics analysis software
developed by Marseille Observatory.} software package was also used.

\clearpage

\section{RESULTS} \label{results}

This section focuses on presenting
(i) optical and near IR images of NGC\,5514;
(ii) 2D spectroscopic evidences for two extra-nuclear
starbursts with OFs and LINER activity;
(iii) a supergiant galactic bubble generated by one of these
starbursts, and an extended complex of H {\sc ii} regions associated
with the other extra-nuclear star formation area; and
(iv) detailed studies of the 2D kinematics and ionized structure.


\subsection{The broad band morphology of NGC\,5514 (optical and near IR images)}
\label{res-morpho}

The CASLEO optical V image (Figure \ref{wholemer}) shows the
whole merger. This system consists of a main body with a radius $r$ $\sim$
35$''$ ($\sim$ 17 kpc), a bright tidal tail with a projected extension of
$\sim$104\farcs8 ($\sim$50 kpc) located to the east of the main body and a
very faint second tail of $\sim$114\farcs4 ($\sim$55 kpc) to the west. We
note that both tidal tails depict a clear or relatively strong continuum
emission mainly in the V and B images.

For the main body, optical and near IR contour images (in
U, B, V, I, J, H and K$_S$; Figures \ref{caimages} and ~\ref{2mimages}) show
the following main structures  (at a resolution of $\sim$1\farcs5--2\farcs0;
FWHM): a main nucleus which is bright in all the observed optical and near
IR broad band images; plus a second nucleus to the SE from the main nucleus
and at a distance of $\sim$ 5.4 kpc ($\sim$11\farcs7), which is bright in
U,  I, and in the near IR bands.
In general, these optical and near IR images depict the typical
features of a pre-merger system.

In addition, several  interesting and unusual features were also detected
in these images. In particular:

\begin{enumerate}

\item
The Palomar deep 103a-O image (broad band B; Fig.~\ref{palimage})
shows several faint
`radial' filaments in the outer regions.
More specifically, on the  north-west side of the main body
at least three radial filaments were found, each one with an
extension of 5\farcs8 ($\sim$2.8 kpc; Figure \ref{palimage}).
Similar types of filaments in emission (and/or in absorption) were
detected in the OF external regions of  NGC\,3256, NGC\,2623,
M\,82, NGC\,3079 and other galaxies (e.g. L\'{\i}pari et al. 2000,
2004a; Cecil et al. 2001).

\item
An interesting feature was found in the (B $-$ I) colour image.
Figure \ref{redening} shows one or two very extended bands through
almost the entire main body of the merger.
These bands  depict strong emission in V and B images.

Since these two bands or structures show similar photometric properties
that the tidal tails, and they  are aligned with the bright east tail,
a simple explanation--for these structures--is that they are
the beginning of the east tidal tail. Which probably emerges from
the distorted spiral disc of the main galaxy that collided.
Thus, the total projected extension of the east tails is  $\sim$175$''$
($\sim$83 kpc).

A second explanation, for these structures, is that they are associated
with the presence of a very extended disc or ring of dust, around the
main nucleus. This type of structures have been detected in several
spiral galaxies (e.g. Feinstein et al. 1990).

\item
In addition, the V and B broad band images show for the second
nucleus very weak continuum emission.
However, in the U, I and near IR bands this nucleus depicts a bright and
compact shape (Figures \ref{caimages} and \ref{2mimages}).
A simple explanation for these observed properties is that they could be due
to the absorption of the emission in the beginning of the tidal tail or
in the possible ring/disk of dust.

\item
An interesting structure in the bright eastern tail is a faint filament extending
at right angles, to the south (already detected by FL88, in their B image:
figure 2a).
It is not easy to find a simple explanation for this structure .

\end{enumerate}

The main galaxy shows the major
photometric axis aligned approximately with the east-west direction.
This galaxy, in the broad bands I and K$_S$, shows a projected diameter
d $\sim$ 50$''$ ($\sim$24 kpc).
The position of the near IR photometric major axis was analysed in detail
by fitting ellipses to the 2MASS K$_S$ image. We found the following
position angle (PA) for the photometric major axis:
PA$_{Ks}$ = 80$^{\circ}\pm$7$^{\circ}$ for 2 $\leq$ r $\leq$ 6 kpc.
This radial range of radius corresponds to
the regions where the major axis PAs show relatively constant values.
Variations of the photometric major axis PA in the
nuclear and circumnuclear regions (for r $\leq$ 2 kpc) were detected,
which could be due mainly to the presence of dust and  blue structures.

A second detailed study of the K$_S$ luminosity profile was performed
for the northern galaxy.
We fitted ellipses 1 pixel
(1\farcs0) wide, and  the surface brightness
along the ellipse major axis at each radius was extracted.
The obtained profile then corresponds to the surface brightness
across the photometric major axis.
A good fit of the K$_S$ surface brightness with the
r$^{1/4}$ law was obtained in the radial range
1 $\leq$ r $\leq$  12 kpc (see Fig.~\ref{perfill}a).
If we combine the  r$^{1/4}$ and an exponential law, a small improvement of
the fit is obtained.
Therefore, for the main galaxy of NGC\,5514, the
near IR (K$_S$ band) photometric profile
corresponds mainly to a bulge dominat disc system; i.e., Sa or Sb galaxy.

The projected diameter of the south-east galaxy is d $\sim$ 30$''$
($\sim$15 kpc), in the  I and K$_S$ broad bands.
A similar study to that performed for the main galaxy was carried out for the
south-eastern galaxy.
We found for near IR photometric major axis (fitting ellipses to the K$_S$
image)  a PA$_{K}$ = 11$^{\circ}\pm$8$^{\circ}$
for 2 $\leq$ r $\leq$ 4 kpc.
A good fit of the K$_S$ surface brightness with the
r$^{1/4}$ law was obtained in the radial range
1 $\leq$ r $\leq$ 7 kpc  (see Figure \ref{perfill}b).
For this galaxy the S/N ratio of the K$_S$ image allows us to fit only
the main component (i.e. the r$^{1/4}$ law). However, a disc
component could be present but at the level of the noise.
Therefore, for the south-eastern galaxy of NGC\,5514 the
near IR  photometric profile
correspond mainly to a bulge dominat system (Sa, Sb, or S0 galaxy).

A detailed study of the K$_S$ and I luminosity profiles in the overlapping
area of the two galaxies suggests that the eastern part of the main system
is positioned (from our point of view) behind the second galaxy.

\clearpage

\subsection{Long-slit spectrophotometry}
\label{res-ls}

Long-slit
spectra of NGC\,5514 were obtained from our {\it Survey of Warm IRAS Galaxies Candidate} .
Using these spectra, we detected (and `re-discovered'; see  FL88)
two extended regions with strong `extra-nuclear' emission lines.
This set of 1D data was the basis for the next 2D spectroscopic study.
In addition, these long-slit observations allow us to study a wide
wavelength range ($\lambda\lambda$ 3800--7500 \AA) and to analyse part
of this wavelength range at high resolution (50 km s$^{-1}$).

On the other hand, it is useful to compare 1D and 2D data, since this
analysis will help us to understand observations of more distant IR
mergers/QSOs, where only 1D observations are available and it is not
possible to separate spatially the different galactic components.

\subsubsection{Moderate resolution spectra}
\label{res-ls1}
\vspace{5mm}

In this section we shall study the following three regions:
a) the main nucleus;
b) a strong extra-nuclear H$\alpha$ + [N {\sc ii}] emission area,
located to the west of the main nucleus, and
c) the second nucleus plus the close extended H {\sc ii} region area,
positioned to the south--east of the main body.

Fig.~\ref{1dspec1}  shows the 1D long-slit spectra obtained with
moderate resolution ($\sim$290--300 km~s$^{-1}$ FWHM) for these
three main regions. Tables~\ref{flux1d} and ~\ref{elr1d}  show
the calibrated fluxes and emission line ratios of these spectra.

The main conclusions from these data are the following:
i) the emission line ratios of the main nucleus are consistent with a
weak LINER;
ii) the spectrum of the western area shows properties of a strong LINER
system, with a  faint continuum;
iii) the second nucleus and the extended H {\sc ii} complex
depict emission line ratios typical of H {\sc ii} regions and `transition'
LINER/H {\sc ii} objects.

In addition, the main nucleus shows the standard absorption
lines of an old stellar population. In particular,  for
Mg {\sc i}$\lambda$5175, (using the $\lambda\lambda$5156--5196 window)
an equivalent width EW $\sim$7.0 \AA\, was measured.

For the western area with strong emission,
the values of the fluxes (presented in Table~\ref{flux1d})
show that this region is the most luminous in [N {\sc ii}] emission line.
For the second nucleus plus the H {\sc ii} regions,
the CTIO and CASLEO spectra (which cover a wide wavelength range)
depict a relatively intense blue continuum; probably associated with the
presence of a hot young massive stellar population.
In addition, Table~\ref{flux1d} shows that this region is the more
luminous in H$\alpha$ emission.

\subsubsection{High  resolution spectra}
\label{res-ls2}

We also studied the three main regions of the main body of NGC 5514 with
CASLEO high spectral resolution spectroscopy (of 50 km~s$^{-1}$).
Figure \ref{1dspec2},
and tables \ref{flux1d} and ~\ref{elr1d} present the spectra,
the values of the fluxes and the emission line ratios.
At this resolution, the presence of multi emission line components (ELC),
in the west/LINER region,  were clearly detected.

In addition, Table~\ref{flux1d} reveals an interesting result:
the two types of long-slit spectra (with moderate and high resolution),
give a value of the excess E(B$-$V)$_{I} =$ 0.0, for the  western area.

Finally, it is important to note that the results obtained from these
1D spectroscopic observations and the optical and near
IR images data (obtained mainly at CASLEO),
were used as the base for the 2D spectroscopic study.
This 2D study was performed mainly in order to clarify  the
nature of the two strong extra-nuclear emission line regions;
and  the physical conditions in both nuclei.

\clearpage

\subsection{Mapping two extra-nuclear starbursts and the nucleus region
(2D spectroscopy)}
\label{res-enl}

Figures \ref{whtmosaic12}(a) and (b) show the red continuum and
the  H$\alpha$ + [N {\sc ii}] emission line maps for the central region
of the main body of NGC\,5514. These images are mosaics obtained by
combining  three individual WHT + INTEGRAL frame of 16\farcs4$\times$12\farcs3.
Each mosaic covers a total area of $\sim$30$''\times$20$''$
($\sim$14.3 kpc $\times$ 9.5 kpc).

FITS maps were obtained from the original ASCII data (following the
procedure described in detail in Section 2.6), with a spatial sampling of
0\farcs9$\times$0\farcs9 and a typically seeing of $\sim$1$''$, FWHM.

Figure \ref{whtmosaic12}(a) shows mainly the continuum emission from the
two nuclei, whereas
Figure \ref{whtmosaic12}(b) depicts the strong H$\alpha$ + [N {\sc ii}]
emission line, mainly from the main nucleus and the two extra-nuclear
extended structures located to the west and east of the main body.
These regions are analysed in detail in the next subsections.

\subsubsection{Mapping the supergiant galactic bubble}
\label{res-bubble}

In order to study the strong extra-nuclear area where we previously detected
extended LINER activity (with long-slit spectroscopy),
the first  WHT + INTEGRAL observation was centred on the
west side of the main body of the merger.
The main nucleus was thus positioned on the left border of this
field (7\farcs9 E and 1\farcs2 N, from the centre of the fibre bundle).

Figure \ref{idab} shows a supergiant galactic bubble, from the
H$\alpha$ + [N {\sc ii}] emission line map, superposed on the
individual fibre spectra (for this INTEGRAL position 1).
Even in this figure alone the relation between the changes in the
the profile of the lines and the structures in the bubbles is obvious.
In particular, in the main knots the spectra show clear broad emission
associated with the presence of multi-components (for details see
Section 3.4), plus very strong [N {\sc ii}]/H$\alpha$ ratios.

Figure \ref{bubblei} presents the red continuum (adjacent to  H$\alpha$)
plus the individual  H$\alpha$,
[N {\sc ii}]$\lambda$6583, [S {\sc ii}]$\lambda\lambda$6717 + 31,
 [O {\sc i}]$\lambda$6300 and [O {\sc iii}]$\lambda$5007 emission line
contour maps for the position 1 (obtained from the 2D spectroscopic data).
The continuum contour map depicts emission only from the main nucleus.
However, all the emission line contour maps show strong extra-nuclear
emission from a supergiant galactic bubble with four main knots.
This bubble shows a distorted ellipsoidal morphology with an elongation  at
PA$_{bubble}$ $=$ 120$^{\circ}\pm10^{\circ}$, with major and minor axes of
13\farcs6 ($\sim$6.5 kpc) and 9\farcs6 ($\sim$4.5 kpc), respectively.
The centre of the galactic bubble is positioned at 8\farcs5 ($\sim$4.1 kpc)
to the west of the main nucleus.

In general, the morphology is  the same for the stronger optical
emission lines (H$\alpha$, [N {\sc ii}]$\lambda$6583,
[S {\sc ii}]$\lambda$6717 + 31). In addition,
the weaker emission lines, [O {\sc i}]$\lambda$6300
and [O {\sc iii}]$\lambda$5007, give maps with relatively similar structures,
but at low S/N (Figures \ref{bubblei}e, f).
However, Figure \ref{bubblei}(c) depicts an interesting difference:
the [S {\sc ii}]$\lambda\lambda$6717 + 31 emission map shows
wide filaments at the border of the bubble, probably associated
with OF structures.
Similar results were found in the mergers with OF NGC 2623 and 3256
(L\'{\i}pari et al. 2004a, 2000). This point will be analysed in Section 4.4.

It is important to study in detail the four main knots detected in this
supergiant galactic bubble with 2D spectroscopic data.
In Fig.~\ref{bubblei}(b) the positions of these knots are depicted and labelled.
We note that different emission line maps show substructures in
knots 2 and 3. In these knots we call the main substructure as `a'
and the second one `b'.
Tables \ref{flux2dt}, \ref{elr2dt}, \ref{flux2df}
and \ref{elr2df} present flux and emission line ratio values for
each main structure in NGC 5514 (including the knots in the bubble),
and also for the  fibres located at the centre of these structures.
For the case of the bubble region,
Fig. 10 shows the positions and numbers for all the fibres in an INTEGRAL
field.

We shall first analyse the properties of the knots using the
emission line maps.
In all these maps (Fig.~\ref{bubblei}) strong emission at the position
of the knots was detected.
More specifically,

\begin{enumerate}

\item
Knot 1, with an elongated shape (at PA$_{K1}$ $=$ 110$^{\circ}\pm10^{\circ}$)
and a major axis of 4\farcs6 $\sim$2.3 kpc,
is the most extended and luminous structure in the bubble.
This knot is the only one with similar strong and  extended morphology in
all the observed emission lines.

For this knot a value of H$\alpha$ flux F$_{H\alpha} =$ 2.5 $\times 10^{-14}$
erg cm$^{-2}$s$^{-1}$ was measured; and the
corresponding luminosity is L$_{H\alpha} = 3.0 \times 10^{40}$ erg s$^{-1}$.
In addition, knot 1 is the main structure closer to
the centre of the bubble.  This extended knot is then a good candidate to
harbour an association of `super massive' star clusters (SSCs) that generate
the OF process and the bubble.

\item
Knots 2, 3 and 4 are less extended  and more compact than  knot 1,
with  diameters of $\sim$2\farcs0 ($\sim$1.0 kpc). These structures are
located close to the border of the bubble.

We note that the emission line contour maps  depict  sub-structures,
especially inside knot 2. This
knot 2  shows strong emission especially in the [S {\sc ii}] line,
whereas knots 3 and 4 show strong emission in the
[N {\sc ii}] and  H$\alpha$ lines, respectively.

\item
The [S {\sc ii}] contour map (Fig.~\ref{bubblei}d) shows clearly at least
four wide filaments emerging from the knots.
Furthermore,
in section 3.5 we present kinematics evidence that these three
`external' knots are probably associated with the areas of
ejection of the ionized gas and rupture of the external shell of the bubble.

\end{enumerate}

The 2D WHT + INTEGRAL spectra of these knots will now be analysed.
Figure \ref{bubbles} depicts the blue and red 2D spectra of these
four main knots. Tables \ref{flux2dt}, \ref{elr2dt}, \ref{flux2df}
and \ref{elr2df}  show the values of
the fluxes, EW, FWHM and luminosities of the emission
lines, and their ratios, for the bubble, the nuclei and selected area
of the complex of H {\sc ii} regions.
In these tables the results of Section 3.4 (about the 
analysis of the multiple emission line systems, in the 2D spectra),
were used.

Tables \ref{flux2dt}, \ref{elr2dt}, \ref{flux2df} and \ref{elr2df}
reveal an interesting fact: inside the bubble,
the spectra for all the main knots depict  LINER characteristics.
These `extra-nuclear' LINER properties could be associated mainly with the
OF process. For a detailed study and discussion of this
point see Sections 3.6 and 4.3.

Another interesting result obtained from these spectra
(Figure \ref{bubbles} and  Table \ref{flux2dt})
is the detection in the knots 1, 2a, 2b, 3a and 3b of Wolf Rayet (WR)
features at $\lambda$4560 \AA. It is interesting to note that
the [N {\sc i}]$\lambda$5199 emission line was also found
in these knots (where we detected the WR bump).
A similar behaviour (i.e. strong WR and [N {\sc i}] emission)
was observed in the spectra of the prototype WR galaxy NGC 6754
(Osterbrock \& Cohen 1982).
These clear WR features suggest the presence of a high number of massive
stars (probably in `young' SSCs), with  ages
t $<$ 6--8 $\times$ 10$^6$ yrs in these knots.

For the main emission line component of the bubble, a flux of
F$_{H\alpha} = 5.5 \times 10^{-14}$ erg cm$^{-2}$ s$^{-1}$ was measured
and a total value of H$\alpha$ luminosity,
L$_{H\alpha} = 6.3 \times 10^{40}$ erg s$^{-1}$, was derived.
The dynamical and physical properties of this young OF in the bubble phase,
will be discussed in Section 4.4.

\subsubsection{Mapping the nuclear region}
\label{res-nucr}

The second position of the WHT + INTEGRAL fibre system was selected in order
to study the two nuclear areas. The  fibre system  was centred
midway between the two nuclei.

Figure \ref{nucleii} shows the continuum adjacent to H$\alpha$,
plus the individual  H$\alpha$, [N {\sc ii}]$\lambda$6583 and
[S {\sc ii}]$\lambda\lambda$6717 + 31 emission line contour maps
(for the position 2).
The continuum contour map depicts mainly the emission from the nuclei.
The emission line  maps show the main nucleus and  the second
nucleus superposed with the extended complex of
giant H {\sc ii} regions.

Figure \ref{nucleis} presents the blue and red  2D spectra for
the two nuclei.
These spectra exhibit  the typical LINER features for the main nucleus
and transition LINER/H {\sc ii} region properties for the second one.

For the position 2, the INTEGRAL blue continuum map clearly
shows the presence of the extended bands/fringes and a faint continuum
from the second nucleus, confirming  the results
obtained in Section 3.1.

\subsubsection{Mapping the extended complex of giant H {\sc ii} regions}
\label{res-h2r}

The INTEGRAL frame at position 3 was centred close to
the complex of giant H {\sc ii} regions (and thus close to
the second nucleus, which is located 3\farcs7 E and
0\farcs2 S from the centre of the fibre bundle). In this area and
using long-slit spectroscopy, we had previously detected
a very extended emission, with
a composite or transition LINER/H {\sc ii} region spectrum.
We note, that objects with transition LINER/H {\sc ii} region
properties are those with intermediate emission line ratios,  
between pure LINERs  and typical H {\sc ii} regions.
These objects would be classified as LINERs except--mainly--that their
[O {\sc i}] $\lambda$6300 line strengths are too small in comparison
with other lines to meet the formal LINER criteria.
For more detailed definitions of these types of activity see 
Barth \& Shields (2000); Ho (1996); Ho, Filippenko \& Sargent (1993, 1997);
Heckman (1996, 1987, 1980).

Figure \ref{h2i} shows the continuum adjacent to H$\alpha$ and
the  H$\alpha$, [N {\sc ii}] and [S {\sc ii}] emission line contour maps
for the position 3.
The continuum contour map shows mainly the emission from the second nucleus.
The emission line contour maps depict an extended H {\sc ii}
region complex, which at H$\alpha$,
[N {\sc ii}]$\lambda$6583 and [S {\sc ii}]$\lambda\lambda$6717 + 31 shows
a compact area of strong emission
line, with the peak  located at 1\farcs7 $\sim$810 pc to
the east of the second nucleus.
In addition, this extended complex shows 
an elongated shape (at PA $\sim$ 20$^{\circ}\pm8^{\circ}$), with 
major and minor axes of 10\farcs8 $\sim$5.1 kpc and 6\farcs0 $\sim$2.9 kpc,
respectively.
The morphology is similar for these three main optical emission
lines  (Figures \ref{h2i} b, c, and d).

Figure \ref{h2s} depicts the blue and red 2D spectra of
selected areas of this H {\sc ii} region complex: the main central
knot and two areas at the border.
Tables~\ref{flux2dt}, ~\ref{elr2dt}, ~\ref{flux2df} and  ~\ref{elr2df}
show the values of the fluxes, EW, FWHM, luminosities of the emission line
and their ratios for these selected regions.
In particular, these plots and tables show:
i) inside of this complex, the values of emission line ratio (ELR) are
consistent with gas ionized by hot stars (i.e. typical of H {\sc ii} regions);
ii) for the second nucleus, the values of ELR (Table~\ref{elr2dt})
are consistent with transition LINER/H {\sc ii} properties,
for an area with radius {\it r} $\sim$ 2$''$ $\sim$ 950 pc;
however the ELR  for the central  fibre ({\it r} $\sim$ 1$''$ $\sim$ 470 pc;
Table~\ref{elr2df}) show LINER properties;
iii) at the border of this extended complex the spectra depict weak but
clear OF component (Figs.~\ref{h2s}c,d) and their values of ELR are
consistent with LINER properties (similar to those observed in the OF of the
galactic bubble).

For this region, a `total'  H$\alpha$ flux of
F$_{H\alpha} =$ 5.9 $\times 10^{-14}$ erg cm$^{-2}$s$^{-1}$ was measured;
and the corresponding luminosity is
L$_{H\alpha} = 6.6 \times 10^{40}$ erg s$^{-1}$. This last value is also
compatible with starburst areas (Kennicutt, Keel \& Blaha 1989).

\clearpage

\subsection{Multiple components in the emission lines
(2D and 1D high resolution spectroscopy)}
\label{res-mcel}

An important point in the study of spectra obtained with relatively high
spectral resolution is the analysis of multiple components (in each emission
line, especially in the stronger lines [N {\sc ii}]$\lambda$6583 and H$\alpha$).
This type of detailed study was performed for several nearby galaxies,
including systems with GW, e.g. NGC 3079, 3256, 1052, 4550, 7332,
Cen A, etc (e.g. Plana \& Boulesteix 1996; L\'{\i}pari et al. 2000; Veilleux et al.
1994; Bland, Taylor, \& Atherton 1987).

For NGC\,5514, 1D and 2D spectra with relatively high resolution
($\sim$50 and 100 km s$^{-1}$) were obtained. From these data 
a detailed study of multiple ELCs was performed (especially
in order to analyse OF motion), and
the following main results were obtained:

\begin{enumerate}

\item
{\it Main Component (MC)}:
 in the two nuclei and in the H {\sc ii} region complex mainly a single
strong emission line component was detected, in $\sim$81 per cent
of the observed field.
This ELC was measured and deblended using the software SPLOT (see section
2.6).

\item
{\it Blue and Red Bubble Component}:
the presence of two strong components were observed,
in the region of the bubble.  These ELCs are
blue and redshifted, in relation to the systemic velocity (of the merger);
and they  were deblended using the software SPECFIT.

\begin{itemize}

\item
{\it Blue Bubble Component (BBC)}:
 the strong blue component was detected in almost the entire extension of
the bubble (which is $\sim$12 per cent of the observed mosaic),
while the red one was observed mainly in the central
and in the knots areas.

Thus, in several regions this blue  ELC were measured blended mainly with
a weak red bubble ELC.
However, in several areas (mainly in the knots) these blue and red bubble
ELCs show similar strengths, so the deblending was more complex.

\item
{\it Red Bubble Component (RBC)}:
the red bubble ELC was detected in $\sim$90 per cent of the bubble
with a very weak strength. Only in the knots of the bubble this
RBC shows strong features. Thus, in order to measure this component
was required a detailed analysis.

However, in the H$\alpha$ + [N {\sc ii}] blend this red ELC
was relatively easily measured, since the emission line
[N {\sc ii}]$\lambda$6583 (from the red ELC) is located
 on the red boundary of this blend.

Figure \ref{mcomp1} shows examples of each component:
MC, BBC, and RBC; in the spectra obtained from the extraction of individual
fibres (for different regions of the main body).

Figure \ref{mcomp2} depicts the emission line map for the red bubble component,
 in the  [N {\sc ii}]$\lambda$6583 line (at the position 1). This plot shows:
1) two strong and compact peaks of emission with a diameter
$d =$ 4$'' (\sim$1.9 kpc) one located close to the position of knot 2
and the other between knots 1 and 4;
2) a weak compact peak located to the north-west of knot 1;
3) a very interesting extended structure detected in the
south-east region of the map (i.e. on the eastern border of the bubble),
with a quadrupolar substructure close to knot 2a.
This extended + quadrupolar structure is probably a main ejection from
the bubble.
This ejection is also suggested by four kinematics velocity field maps, and
the [S {\sc ii}] emission image (see Sections 3.5 and 3.3).
More specifically, this RBC in the extended ejection area shows:
1) an elongated shape with an extension of 8\farcs3 $\sim$4.0 kpc;
2) an alignment with the major axis of the bubble
(PA$_{ejection} =$ 100$^{\circ}\pm10^{\circ}$).
We note that in this ejection area the red bubble system is the only
component detected.

\end{itemize}

\item
{\it Weak Outflow Component (OF1)}:
in the complex of H {\sc ii} regions several areas show
weak blue outflow components
in addition to the main single systems.
In Section 3.3 the plots of this OF component were presented
(Figs.~\ref{h2s}c,d)

\end{enumerate}

It is important to remark that
for the nearest supergiant galactic bubble in NGC 3079, Veilleux et al.
(1994) already reported  similar results: the presence of two clear emission
line systems, also blue and redshifted from the V$_{sys}$.
They suggested--for NGC 3079--that these two ELCs are associated
with the emission from the external shell of the bubble. Thus, following
this last suggestion (Veilleux et al. 1994 and Filipenko \& Sargent 1992),
we propose for the blue and red ELCs (in the bubble of NGC\,5514) a similar
interpretation: that the BBC come from the nearest side of the
external shell, and probably for this reason this blue ELC was observed in
almost all the extension of the bubble.
Meanwhile, the RBC came from the far side, and thus it is partially
obscured in several areas. Furthermore, probably we are looking the red ELC
through an optically thin bubble, which is consistent with the 1D  and 2D
spectroscopic results, in the sense that the internal reddening in the
bubble shows mainly low values of the excess $E(B-V)_{I}$ (in the range
0.0--0.5, see Tables~\ref{flux1d}, 4 and 6).

In conclusion, with the spectral resolution of this study we can identify
at least 4 different emission line  systems. These results are specially
important for the generation and interpretation of the
velocity fields and emission line ratios maps.

\clearpage

\subsection{Mapping the ionized gas kinematics (2D spectroscopy)}
\label{res-kin}

The ionized gas velocity field (VF) maps were constructed on the basis of the
result of the  previous analysis of multiple component in the emission lines
(Section 3.4).
Specifically,  for each of the main 4 ELCs
the centroid velocities, fluxes and FWHM were measured and deblended
(for the lines [N {\sc ii}]$\lambda$6583, H$\alpha$,
[S {\sc ii}]$\lambda$6731, [O {\sc i}]$\lambda$6300,
[O {\sc iii}]$\lambda$5007 and H$\beta$),
using the SPECFIT and SPLOT software tasks.

Considering that more than 65$\%$ of our emission line measurements show
values in the range 4 $>$ [N {\sc ii}]\,$\lambda$6583/H$\alpha >$ 1,
we present and discuss mainly the [N {\sc ii}]\,$\lambda$6583
velocity field map, which has lower uncertainties and a better
resolution than the H$\alpha$ radial velocity map.


Fig.~\ref{whtmosaic34} shows a mosaic of the 3 observed INTEGRAL 
[N {\sc ii}]$\lambda$6583 velocity field for the main body of NGC\,5514.
This mosaic covers an area of $\sim$30$''\times$20$''$
(14.3 kpc$\times$9.5 kpc).
The uncertainties vary from approximately $\pm10\,$km~s$^{-1}$ in the
regions where the emission lines are strong (mainly the nuclei and the two
extra-nuclear starburst), to $\pm$30 km~s$^{-1}$ for the weakest lines
away from the central area.

Following the procedure performed previously for NGC 3256 and 2623
(Lipari et al. 2000, 2004a),
the kinematics maps were constructed mainly for the main/strong ELC:
i.e., the blue bubble system (inside the bubble), the red bubble component
(in the knots 2, 3, and 4, and in the south-east areas
of the shell), and the single main body component in the remaining areas.

In general, the colour [N {\sc ii}]$\lambda$6583 emission line VF map
(Fig.~\ref{whtmosaic34}) shows:

(i) an extended lobe of blue/approaching velocities in the region of the
bubble, from which an extended ejection emerge;
(ii) an extended area of strong red/recessing velocities to east border of our
mosaic, including the region of the second nucleus;
(iii) a circular lobe of blueshifted  velocities, which is located
between the  extended blue and red lobes;
(iv) a small lobe of blueshifted  velocities located in the north-east border of
the mosaic field.

Using all the observed 2D [N {\sc ii}]$\lambda$6583 and H$\alpha$ emission
lines, a mean value of the systemic velocity
$V_{Sys}$ = 7358$\pm20$ km s$^{-1}$ was measured, while the
main and secondary nucleus have  radial velocities of
$V_{MainNuc}$ = 7280$\pm20$\,km\,s$^{-1}$ and
$V_{SecNuc}$ = 7680$\pm20$\,km\,s$^{-1}$. This  $V_{Sys}$ was used as the
zero value for all the velocity fields.

\subsubsection{The kinematics of the supergiant galactic bubble}
\label{k-bubble}

Fig.~\ref{kvfmapb} displays detailed 2D kinematics contour maps
 for the position of the bubble (position 1; 16\farcs4 $\times$ 12\farcs3),
which were obtained from the measured emission lines
[N {\sc ii}]$\lambda$6583, H$\alpha$ and [O {\sc iii}]$\lambda$5007.

In the study of the general properties of the VF (in the bubble region) the
following main results were found:

\begin{enumerate}

\item
Probably, the most interesting feature for this area (and for all the VF)
is the clear extended  ejection of $\sim$ 4 kpc, in the east side of the
bubble, which started in a position close to the knot 2 and is aligned
with the major axis of the bubble (at PA$_{ejection}$ $=$
100$^{\circ}\pm10^{\circ}$ and PA$_{bubble}$ $=$  120$^{\circ}\pm10^{\circ}$).

In this merger the [O {\sc iii}]$\lambda$5007
emission line is very weak (see Section 3.3). However, even in this case the
VF map shows in its south east border of the map, the feature associated with
the main ejection process.

In addition, this ejection shows   very interesting substructures inside.
In particular, at small scale a quadrupolar  OF/ejection
was found (for details see the last paragraphs in this
section).

\item
In the west side of the bubble (relatively close to the knot 1)
there is an approaching/blueshifted velocities
region, which is probably an ejection, also aligned with the PA of the
major axis of the bubble. We note that two opposite ejections in the
extreme of the ellipsoidal bubble are expected in the OF rupture phase.

This blueshifted ejection is clearly weaker than that observed
in the east side. However, this feature is located close to
the west border of our observed mosaic.

\item
Another two weak ejections were found, associated with the knots
3 and 4. We note that these features were observed in more detail in the
[S {\sc ii}] VF (see the next subsection).

\item
In this supergiant galactic bubble--close to the centre--the
following mean OF  velocities were measured, for the blue and
red components:
$V_{OF blue}=(-320\pm20)$\,km\,s$^{-1}$ and
$V_{OF red}=(+265\pm25)$\,km\,s$^{-1}$,
relative to the systemic velocity.

\end{enumerate}

Fig.~\ref{kvperfilv} shows the [N\,{\sc ii}]$\lambda$6583 velocity profile,
obtained from the velocity field and through the main ejection
(at PA $\sim$ 100$^{\circ}$).
This plot exhibits a variation of velocity in the region of the knot 2 (i.e.,
the beginning of the main ejection)
of $\Delta$V $\sim$ 1000 km s$^{-1}$. This high value in the variation of
velocity could be  explained as the wall/region of giant OF shocks.
We note that the other possible process that could explain detected
$\Delta$V in the VF of merger (such as tidal disruption of a disk, etc)
shows mostly very low velocity values, as it is observed in the velocity field
of nearby mergers ($\Delta$V$_{max}$  $\leq$ 300 km\,s$^{-1}$, e.g.
 L\'{\i}pari et al.\ 2000, 2004a).
It is important to remark that in order to confirm this result an interesting
test was performed: we derived a second
[N\,{\sc ii}] velocity field (for this position 1) following the blue
bubble component, even where this BBC is very weak. Thus, the red
component was included only where the BBC disappear.
Even in this case  a similar plot to Fig.~\ref{kvfmapb} was obtained and
only the region of high $\Delta$V shows a
small shift in position.

In order to analyse in detail the sub-structures in the ejections and the
knots, a basic study was performed, following the good result obtained
using the filtering technique in the study of the VF in NGC 2623.
We made probably the simplest
approach to the mean motion, using a Gaussian filtering, which gives a
smooth [N {\sc ii}]\,$\lambda$6583 velocity field 
showing the 4  main ejections more clearly (Fig.~\ref{smooth}a). Furthermore,
at the beginning of these 4 ejections  very interesting structures in
the residuals map were found.

Specifically, after the subtraction of this smooth VF,
the residuals show: first a  `quadrupolar' structure
(Fig.~\ref{smooth}b), in the beginning of the main ejection, close to
the knot 2.
This quadrupolar OF feature shows  an `X' shape with redshifted velocity
values, plus 3 or 4 blueshifted symmetric lobes
(inside of this red `X' structure).
The residual in this area reaches values close to
$\vert$V$_{max}$$\vert \sim$ 400 km s$^{-1}$.
We note that
this OF quadrupolar structure is similar to those detected in the  OF of
some planetary nebula (Manchado, Stanghellini, \& Guerrero 1996; Gerrero \&
Manchado 1998; Lopez et al. 1998; Muthu \& Anandarao 2001).

Two similar residual OF structures were detected close to the knots 3 and 4,
also at the beginning of the other two ejections. However, in these cases the
velocity residuals show `bipolar' shape, with  redshifted velocity
values, plus two blueshifted symmetric lobes (in each area;
Fig.~\ref{smooth}b).
The values of the residual--in these knots region--reach
$\vert$V$_{max}$$\vert \sim$ 300 and 250 km s$^{-1}$, respectively.

Another residuals structure of approaching/blueshifted velocity
was detected, at the west side of the bubble and relatively close
to the knot 1 (Figs.~\ref{smooth}a, b).
This structure is probably part of a blueshifted ejection, also aligned with
the PA of the major axis of the bubble.
The velocity values of these blueshifted residuals reach  a
V$_{max} \sim$ --280 km s$^{-1}$ (which is also a very high value).

We remark that only for this INTEGRAL Position 1 the values of the residuals
(obtained from the subtraction of a smooth VF) show $\vert$V$\vert \geq$ 100
km s$^{-1}$. The residuals structures detected in
the bubble/position-1 have velocity values  $\vert$V$\vert \geq$ 250 km
s$^{-1}$, confirming the real nature of these features,
which are also evident in the original VF map.

In general, these kinematics results are consistent with
expansion of the shell and with the beginning of the blowout GW phase.
A discussion of this process is presented in Section 4.4.

\subsubsection{The kinematics of the bubble in [S {\sc ii}]}
\label{k-bubbles2}

In Section 3.3 we found that in the bubble region the
[S {\sc ii}] emission line map shows clear extension/wide--filaments,
which start in the 3
more external knots. Therefore it is important to study in detail the
VF of this emission line. This blend of
[S {\sc ii}]$\lambda\lambda$6716 + 31 is strongly absorbed by our
atmosphere, in the blue side. However, at the red border
of this [S {\sc ii}]  blend, the line [S {\sc ii}]$\lambda$6731  is
located far from the absorption range; and thus this line  was measured easily.

In addition,
we note that for this [S {\sc ii}]$\lambda$6731 emission line, mainly the
main component was measured, since this [S {\sc ii}]
blend was absorbed in the blue side (which preclude a clear separation
between the BBC and RBC).
Fig.~\ref{kvfmapb}b shows the VF contour map for the main component of
the line [S {\sc ii}]$\lambda$6731.
Which is practically a VF for the BBC in 90 per cent of the bubble, but
in the area of the 3 external knots  the RBC is the
dominant component.

This plot shows probably the most clear features of the 3 main
ejections. In particular:

\begin{enumerate}

\item
Two weak ejections were found, associated with the knots 3 and 4.
These  ejections are almost perpendicular and less extended
than the stronger one, reaching $\sim$1--2 kpc.
We note that the main ejections start in one of the 3
`external' knots.
Probably, these results are consistent with the fact that the bubble is in
the beginning of the rupture phase.

\item
The main/stronger ejection was also clearly observed in this [S {\sc ii}]
emission line VF.
This feature was detected to have similar characteristics to
the previously presented VFs.

\end{enumerate}

It will be discussed in Section 4.4 a possible explanation of the fact
that  [S {\sc ii}]  is probably one of the best tracers of the OF process.

\subsubsection{The kinematics of the nuclei region}
\label{k-nr}

Fig.~\ref{kvfmapnc}(a) displays detailed 2D kinematics contour map,
for the position of the nuclei (position 2), obtained from
the measured [N {\sc ii}]$\lambda$6583 emission line.
This map was constructed mainly for the main emission line component.

From the study of the general properties of the VF in this  region, we
remark the following main results:
at the north-east border of the field, the isovelocity lines show a
strong stretching (probably associated with a high concentration of mass).
In addition, almost all the area of the main extended ejection
(of the bubble) is included in this field.

In the area of stretching isovelocity lines the
B image shows a faint  knot (which was also observed in the B contour image
published by FL88: their Fig. 2a).
In section 4.1 it will be discussed the possibility that the properties found
in this region  could be associated
to the presence of a satellite of the original galaxies that collide.

\subsubsection{The kinematics of the extended complex of giant H {\sc ii}
region}
\label{k-h2r}

Fig.~\ref{kvfmapnc}(b) displays detailed 2D kinematics contour map,
centred in the giant H {\sc ii} regions  complex  
(position 3), for the measured [N {\sc ii}]$\lambda$6583 emission
line.
This map was constructed partly for the single main component and partly
for the red `bubble' system.

We remark the following properties of the VF (in this area):

\begin{enumerate}

\item
The  H {\sc ii} region complex shows mainly red--shifted velocities.
However,  there is also a very  interesting blue feature:  a `circular'
blue-shifted lobe, which is located  close to the main ejection of
the bubble.
This circular lobe is probably associated with the interaction/superposition
of the two main OF process (associated with the bubble and the
H {\sc ii} complex).

We note that this `circular' structure was also observed in the
ELR map of [S {\sc ii}]${\lambda 6717 + 31}$/H$\alpha$ (see Section 3.6.3),
however this feature was not detected
in the [N {\sc ii}]$\lambda$6583/H$\alpha$ ELR map.
This result confirms that the [S {\sc ii}]  emission line is one of the
best tracers of OF + shock processes.

\item
For the weak OF components detected at the border of this H {\sc ii} regions
complex,  the following mean values of velocities were measured:
$V_{OF blue}=(-240\pm25)$\,km\,s$^{-1}$ and
$V_{OF red}=(+100\pm30)$\,km\,s$^{-1}$,
in relation to the main components.

\end{enumerate}

\clearpage

\subsection{Mapping the emission line ratios and widths
(2D spectroscopy)}
\label{res-elrw}

From  2D spectroscopic data -which covers practically
the main structures of the main body- we measured the
fluxes and FWHM of the ELC in order to investigate the
structure of ionization and physical conditions of the gas.

These emission line ratio maps were constructed using the techniques
described in Section 2.6, and using the main result of the study
of multiple ELC (section 3.4).
In particular, these maps were generated with the same components already
included in the kinematics maps: mainly the blue bubble system (inside
the bubble), the red bubble component (in the knots and the south-east area
of the shell), and the single main body component in the remaining regions.

A first and general study of the emission line ratios  in NGC\,5514,
was performed using the `standard' diagnostic diagrams of Veilleux \&
Osterbrock (1987).
Fig.~\ref{elrdiag} depicts the values of ELRs (obtained from
Tables~\ref{elr2dt} and ~\ref{elr2df}) for the bubble, the nuclei and
selected areas of the complex of H {\sc ii} regions.
These plots show for these structures, mainly LINER and transition
LINER/H {\sc ii} properties. These LINER characteristics
will be studied in detail in the next subsections.

\subsubsection{The emission line ratios and widths of the
supergiant galactic bubble region}
\label{er-bubble}

Fig.~\ref{elrmapbu} displays detailed 2D contour maps for the position 1
(16\farcs4 $\times$ 12\farcs3, with 0\farcs9 spatial sampling) for the ELRs:
[N {\sc ii}]$\lambda$6583/H$\alpha$,
[S {\sc ii}]${\lambda 6717 + 31}$/H$\alpha$,
and the FWHM of the [N {\sc ii}]$\lambda$6583 emission line.

Fig.~\ref{elrmapbu} and Tables~\ref{elr2dt}, ~\ref{elr2df} show
interesting features and data. We remark:

\begin{enumerate}

\item
The [N {\sc ii}]$\lambda$6583/H$\alpha$ 2D map (Fig.~\ref{elrmapbu}a) shows
in 100$\%$ of the galactic bubble very high values: i.e.
[N {\sc ii}]$\lambda$6583/H$\alpha$ $>$ 1. This result, obtained
from 2D map, was verified using the individual spectrum of each fibre.

\item
The [S {\sc ii}]${\lambda 6717 + 31}$/H$\alpha$ 2D map (Fig.~\ref{elrmapbu}b)
also shows in $\sim$100$\%$ of the galactic bubble high values
of the ratio [S {\sc ii}]$\lambda$6717 + 31/H$\alpha$ $>$ 0.8. Thus
confirming that these high values of the [S {\sc ii}]/H$\alpha$ and
[N {\sc ii}]/H$\alpha$ ratios could be associated mainly to the effects
of shock ionization in the out flowing gas (Dopita \& Sutherland 1995;
Heckman 1980, 1996; Heckman et al. 1987, 1990; Dopita 1994).

\item
In the line diagnostic diagram [O {\sc i}]$\lambda$6300/H$\alpha$ vs.
[S {\sc ii}]$\lambda$6717 + 31/H$\alpha$ (Heckman et al. 1990), these
values are also consistent with ionization by shocks heating.

We note that the [O {\sc i}]$\lambda$6300/H$\alpha$ and
[O {\sc iii}]$\lambda$5007/H$\beta$ maps depict mainly a clear lobe
located between the position of the knots 1 and 4.

\item
The FWHM--[N {\sc ii}]$\lambda$6583 2D map shows high values
for the bubble area, in the range of 300--600\,km~s$^{-1}$.

\end{enumerate}

Therefore, in the galactic bubble of NGC\,5514 these INTEGRAL 2D emission
line ratios maps show a very interesting result:  
mainly LINER properties associated with an ionization process produced
 mainly by  shocks heating in a large scale OF event, were detected
 (see Section 4.3).

\subsubsection{The emission line ratios and widths of the nuclei region}
\label{er-nr}

Fig.~\ref{elrmapnu} displays detailed 2D maps (for the position 2),
for the emission line ratios
[N {\sc ii}]$\lambda$6583/H$\alpha$,
[S {\sc ii}]${\lambda 6717 + 31}$/H$\alpha$,
and the FWHM of the [N {\sc ii}]$\lambda$6583 emission line.
These maps were constructed mainly for the main component and
for the red bubble component (in the ejection area).

Fig.~\ref{elrmapnu} and Tables~\ref{elr2dt}, ~\ref{elr2df} show the
following main features and data:

\begin{enumerate}

\item
In the main nucleus the
emission line ratios show the following mean values:
[N {\sc ii}]$\lambda$6583/H$\alpha$ = 1.70,
[S {\sc ii}]$\lambda$6717 + 31/H$\alpha$ = 0.95,
[O {\sc i}]$\lambda$6300/H$\alpha$ = 0.71, and
[O {\sc iii}]$\lambda$5007/H$\beta$ = 2.00. In general,
these are typical values of emission line ratios of LINERs associated to
shock heating (Dopita \& Sutherland 1995; Heckman et al. 1990).

In the  [O {\sc i}]$\lambda$6300/H$\alpha$ vs.
[N {\sc ii}]$\lambda$6583/H$\alpha$ diagram, the emission line ratios of
the main nucleus of NGC\,5514 are consistent with LINERs and shock heating
regions (Heckman 1987).

For this nucleus, the FWHM--[N {\sc ii}]$\lambda$6583 2D map shows
values in the range of 230--250\,km~s$^{-1}$.

\item
In the second nucleus the
emission line ratios show the following mean values:
[N {\sc ii}]$\lambda$6583/H$\alpha$ = 0.53,
[S {\sc ii}]$\lambda$6717 + 31/H$\alpha$ = 0.48,
[O {\sc i}]$\lambda$6300/H$\alpha$ = 0.14, and
[O {\sc iii}]$\lambda$5007/H$\beta$ = 0.90.
These values are consistent with those of transition objects,
between LINER and H {\sc ii} regions (Ho 1996; Barth \& Shield 2000).

For the second nucleus, the FWHM--[N {\sc ii}]$\lambda$6583 2D map shows
low values, in the range of 150--180\,km~s$^{-1}$.

\end{enumerate}

Thus, both nuclei show emission line properties and FWHM values consistent
with LINER process associated with OF + shocks events (plus a contribution
of H {\sc ii} regions, in the case of the second nucleus).

\subsubsection{The emission line ratios and widths of the extended complex
of giant H {\sc ii} region}
\label{er-h2r}

Fig.~\ref{elrmapco} displays detailed 2D maps (for the position 3),
for the emission line ratios
[N {\sc ii}]$\lambda$6583/H$\alpha$,
[S {\sc ii}]${\lambda 6717 + 31}$/H$\alpha$,
and the FWHM of the [N {\sc ii}]$\lambda$6583 emission line.
These maps were constructed mainly for the single main body component (and
for the red `bubble' system, in the area of the ejection).

Fig.~\ref{elrmapco} and Tables~\ref{elr2dt}, ~\ref{elr2df} show
interesting features and data; and we remark:

\begin{enumerate}

\item
In this area
the emission line ratios maps show
--for the main ELC--the typical values of giant H{\sc ii} regions
(Terlevich et al. 1991; Kennicutt et al. 1989; and references therein).

\item
Probably the most interesting emission line ratio feature in this area
is the circular structure observed in [S {\sc ii}]/H$\alpha$ associated
with a similar OF structure (detected in the kinematics map of this area,
Section 3.5.3), with a radius r $\sim$ 3 kpc.
We note that again the [S {\sc ii}] and  the [S {\sc ii}]/H$\alpha$ ratio
are the best tracers of OF process (see Section 4.4, for a discussion of
this point).

\item
The area of emission line peak (in the H {\sc ii} complex)
shows emission line ratios consistent with transition
objects between LINER and H {\sc ii} regions (Ho 1996; Barth \& Shield 2000).

This result is similar to that found for the second nucleus, and both are
probably associated with OF process, plus the superposition with H {\sc ii}
regions.

\item
We have already depicted that at the border of this 
H {\sc ii} region complex, the spectra show weak OF components.
These OF and the corresponding main components,
present LINER properties (see Tables 5 and 7).

\item
For this complex of H {\sc ii} regions mainly low values of
FWHM-[N {\sc ii}]$\lambda$6583 were found: in the range $\sim$150--240 km
s$^{-1}$. However, at the border (of this complex, where we found OFs)
the FWHM reach values of 300--400 km s$^{-1}$.

\end{enumerate}

In the next Sections will be discussed mainly the properties of these
extra-nuclear starbursts two areas, with young outflow processes and
LINER and LINER/H {\sc ii} properties (probably generated by the
ongoing merger event).

\clearpage

\section{DISCUSSION} \label{discussion}

In this section  mainly the properties of the pre--merger
process and the two extra-nuclear
starbursts with outflow + supergalactic bubble and LINER activity
will be discussed.
In addition, the probable association between galactic winds and LINERs will
be analysed.
Finally, the properties and the role of the galactic winds
in NGC 5514, IR mergers and IR QSOs, will be considered.

\subsection{The colliding galaxies in NGC\,5514}
\label{dis-coll}

\subsubsection{The pre--merger process in NGC\,5514}
\label{dis-coll1}

Fig.~\ref{mergerc} shows a schematic diagram of the likely orbital
plane geometry, of the NGC 5514 merger; indicating probable location
of our point of view. This diagram was performed according to the
kinematics and morphological results obtained in Sections 3.1 and 3.5.

From Section 3 and FL88, the main properties of the original galaxies
that collide, can be summarized as follows:
(i) the main galaxy has a diameter d$_{1} \sim$50$'' (\sim$24 kpc), mass
M$_{1} \sim 1.8 \times 10^{11} M_{\odot}$, velocity
$V_{1}$ = 7280\,km\,s$^{-1}$, magnitude V = 15.9, colour (B-V) = 1.2,
(V-R) = 0.8, (V-I) = 1.6 (for r $<$ 5");
and (ii) the second galaxy has a d$_{2} \sim$30$'' (\sim$15 kpc),
M$_{2} \sim 0.9 \times 10^{11} M_{\odot}$, $V_{2}$ = 7680\,km\,s$^{-1}$,
V = 17.1 mag, (B-V) = 1.1, (V-R) = 0.8, (V-I) = 1.7.
These values are consistent with those measured in early Sa and Sb
spiral galaxies. We note that in the NASA Extragalactic Database,
the main and the second galaxies are associated with
Sa (NGC 5514--NED1)  and Sb (NGC 5514--NED2) spirals, respectively.

The size of the east tail in NGC 5514 (83 kpc) can give us a first estimation
of the age of the merger.
If an escape velocity of 100 km s$^{-1}$ is assumed for the material in
the tails (Murphy et al. 2001, Colina et al. 1991)  a value of
the age of t$_{merger}$ $\sim$8 $\times$ 10$^{7}$ years was obtained.

A high percentage of the simulation of mergers, were
performed  mainly for collision of equal mass spiral and with disk/halo
components (see for references Barne \& Hernquist 1992, 1996). However,
there are several detailed simulations using different
initial conditions in the merger process (including different values for the
mass ratio). For example, Howard et al. (1993) published a {\it `Simulation
Atlas of Tidal Features in Galaxies'}, with more than 1700 images, obtained
from a wide range of initial conditions.
On the other hand,
Hernquist (1992), Mihos \& Hernquist (1994a,b, 1996), Barne \& Hernquist
(1996) and others already analysed the collisions between spiral with
disk/halo/bulge components. Furthermore, Hernquist (1992) already suggested
that mergers between bulge dominant galaxies are the main type of collisions,
that  more probably could evolve into ellipticals.
Using the atlas of Howard et al. (1993) we found that several types of
unequal mass galaxy collisions could generate a pre--merger morphology
similar to that of NGC 5514, with only one bright tail.
From this comparison and as a new approximation  an age for the merger
t$_{merger}$ $\leq$ 10$^8$ yr (after the first encounter) was derived. Thus,
these results suggest that this merger has an intermediate age.

This  {\it `Simulation Atlas of Tidal Features
in Galaxies'}, also depicts interesting structures inside  the main body
of the mergers, associated with the formation and the beginning of the tidal
tails. In Section 3.1,  this type of structures was probably detected,
superposed to  the main body of this pre--merger.

FL88 suggested that this pre--merger is similar to the `Antennae' and
IRAS 23128--5919: i.e. similar to the collision between two giant
spirals, with two bright tails.
However, the results obtained in Section 3 suggest that the
pre--merger is more similar to those systems that evolve from collision
of bulge dominant galaxies with unequal mass, which generate mainly one
bright tail. In addition,
it is interesting to remark that in our {\it `Survey of Warm IRAS
Galaxies'} we found an IR merger with very similar morphology to that
observed in NGC\,5514. More specifically, in IRAS 03109--5131 (de Grijp et
al. Catalogue N--076, ESO 199--IG 023; z$_{sys}$ = 0.07823,
L$_{IR[8-1000 \mu m]} = 1.0 \times 10^{11} L_{\odot}$, Lipari et al.
1991b) we also found a pre--merger system between two disk galaxies of
 unequal mass (with M$_1$/M$_2$ $\sim$2) that also shows only a bright tail.
This system depicts two compact nucleus of different sizes (R$_1$/R$_2
\sim$2) with a projected separation of d $\sim$ 11 kpc, and both with
Seyfert 2 properties (Lipari et al. 1991b; Fehmers et al. 1994).
Thus, these two Seyfert nuclei could be generated in the early phases of
the merger process (or these AGNs were already formed in the original
galaxies that collided, before the merger event). For NGC 5514, we found
that at least two extra--nuclear starbursts were probably generated
by the  merger event, according mainly to the ages derived for the starbursts
and the merger (see Sections 4.2 and 4.4).
In addition, this proposition is also supported by observational and
theoretical results, which suggest that the merger process frequently
drives large amounts of molecular and ionized gas to the central regions
leading to massive star formation events (see for references Section 1.1).

\subsubsection{A multiple-merger model for  NGC\,5514}
\label{dis-mmm}

Probably, the two spiral + bulge galaxies (that collide)
have satellites, thus the multiple merger scenario needs to be
considered. In order
to explore this type of events it is required to perform detailed
numerical simulations, which are actually in progress (Lipari et al. 2004,
in preparation).

The first results of this kind of simulations
suggest that the two massive spiral galaxies are the main components of the
merger event, and that the satellite generally merges later
(Taniguchi 2004, private communication). This type of composite merger model
(main plus minor mergers) could give a good explanation for the stretching
of the isovelocity lines in the VF, detected at the north--east border of
the field/mosaic  (which is normally indicative of the presence of a
significant concentration of mass).

In NGC 3256, we found that the best model or simulation
that explains the interesting morphology observed (with 3 nuclei connected
with 3 blue asymmetrical spiral arms) is a multiple merger model between two
gas rich massive spiral + bulge galaxies plus a satellite (Lipari et al. 2000,
2004c).

\subsubsection{The NGC\,5514 small group of galaxies}
\label{dis-coll2}

It is important to study the properties of NGC\,5514 in relation to
NGC 5519, since these two systems are very close (in space and velocities)
and could evolve in the future to a multiple merger.
Probably, in the past the two original colliding galaxies (of NGC\,5514)
plus NGC 5519  formed  a small group.

The Sa spiral galaxy NGC 5519 is located at 13.6$' \sim$390 kpc and at PA
130$^{\circ}$, from NGC\,5514. Their heliocentric velocity is V = 7480
$\pm$30 km s$^{-1}$ (FL88). Therefore, NGC 5514 and 5519--with $\Delta$V
= 120 km s$^{-1}$--are probably associated. We note that
Mould et al. (1993, 1995) included NGC 5519 in their study of nearby groups
and clusters of galaxies, within the 100 Mpc.

In NGC 3256 we also found a similar result: i.e.
this merger is part of small group, which could evolve to a multiple
merger (or even to a merger between mergers; L\'{\i}pari et al. 2000).


\subsection{The starbursts in NGC\,5514}
\label{dis-sb}

In Section 3 two extra--nuclear areas with interesting
starburst + OF + LINER properties, plus two nuclei with
LINER plus shock characteristics were found.
Now  these results will be discussed:

\begin{enumerate}

\item
{\it Bubble Region}:
the extra--nuclear supergiant galactic bubble found in NGC 5514 is
probably one of the few young bubble clearly associated
to a starburst, since this type of `supergiant' objects
were previously detected mainly in the dense nuclear
environment and in the post--blowout phase.

Thus the extra--nuclear starburst that originate
 the bubble, probably associated with knot 1, is at least
similar to a powerful  nuclear starburst.
In section 3, for this knot 1 a value of H$\alpha$ luminosity
L$_{H\alpha} = 3.0 \times 10^{40}$ erg s$^{-1}$ was measured.
This luminosity value is consistent with those of strong starburst regions
(Kennicutt, Keel \& Blaha 1989).

The presence of  Wolf Rayet features detected in 3  main
knots of the bubble is clearly indicative of a large number of massive
stars, probably in young SSCs with ages less than 6--8 $\times$ 10$^6$ yr
(see Armus, Heckman, Miley 1988; Conti 1991).
In Table~4,  the fluxes and luminosities of the  WR lines (measured in the
bubble knots) were presented.
This table  shows medium/high WR luminosities values
(between $1.4\times10^{39}$ and $6.5\times10^{39}$ ergs\,s$^{-1}$),
when they are compared with the range found even for WR galaxies (see Armus
et al. 1988, their Table 3; Conti 1991; Schaerer, Contini \& Pindao 1999):
$1.0\times10^{38}$ (for Mrk 724), and $2.9\times10^{41}$
(for IRAS 01003-2238) ergs\,s$^{-1}$. It is interesting to note that
one of the highest value of WR emission line luminosity known was
detected in an extreme velocity OF IR QSO.

In addition, it is important to discuss the general structure of the
bubble, including the knots, in NGC 5514.
The emission line maps show: probably a central SSC
or an association of SSCs, which are surrounded by knots/`arcs' of new star
formation events (at the border of the shocked bubble).
This type of structure
was already observed in several nearby starbursts as `30 Dor, and the
extra--nuclear bubble in NGC 6946' (Meaburn 1980; Redman et al. 2003;
Larsen et al. 2002).
Furthermore, this type of structures was
detected even in starbursts in distant IR QSOS, as Mrk 231, and
IRAS 01003--2238 (Lipari et al. 2003, 1994; Surace et al. 1998).

In general,
we are investigating in IR mergers and IR QSOs---using WHT + INTEGRAL 2D
maps---whether in these shells
giant galactic shocks could generate new star formation episodes by
compressing the ISM (see for references Larsen et al. 2002). These
mechanisms could produce the `dense shell of star-forming knots'
detected in  the arc of Mrk\,231 (a BAL +  Fe\ {\sc ii} IR merger/QSO),
and also the chain/arc of
`extremely blue star-forming knots' in the Wolf--Rayet QSO
IRAS\,01003$-$2238 (both IR QSOs are systems with extreme velocity OF;
L\'{\i}pari et al.\ 2003, 1994).

Furthermore,
at larger scale the presence of chain galaxies at high
redshift was recently explained by mergers of sub galactic gas clumps plus
SN explosions that generate GW + hyper giant-shells (of several hundred kpc),
in which by gravitational instability intense star formation occurs.
Then these systems collapse gravitationally into spheroid systems
(Taniguchi \& Shioya 2001). Thus, this is a similar process to that
proposed for the origin of the external knots in the bubble of NGC 5514,
but at Mpc scale.

\item
{\it Complex of H {\sc ii} Regions}:
in this extended complex  there is clear evidence
that the starburst and the OF process (which probably started to generate
the cavity) are in a young/early state.
More specifically, this complex is probably in the first starburst
phase (0--3 $\times$ 10$^6$ yr), which is dominated by hot main sequence
stars with H{\sc ii} regions spectra. This phase  is also associated to
the presence of dust and IR emission (Terlevich et al. 1993; Franco 2004,
private communication).
This last point is due to the dust present in any star formation region
and also to the large amount of dust synthesized by the most massive stars
(during the $\eta$-Carinae phase before becoming WR star;
Terlevich et al. 1993).

\item
{\it Nuclei Regions}:
since both nuclei depict values in the emission line maps which are consistent
with LINERs associated with OF/sock processes,  their properties
will be discussed in detail in the next section.

\end{enumerate}

The  starburst activity detected in this merger must have occurred
at intermediate age  in the history of the interaction,
since the age  of the pre--merger is $\sim$10$^8$ yr, and the age of
the older extra--nuclear starburst is  6--8 $\times$ 10$^6$ yr.

In the last decades, special attention was devoted to studying the
relation between
starburst and merger processes, mainly in collision of equal mass galaxy (i.e., in
major mergers; Mihos \& Hernquist 1994b, 1996; Barnes \& Hernquist 1996).
However, it is also an important and frequent event the process of starburst
associated with mergers/collisions of unequal mass  galaxy (Bernlohr 1993;
Mihos \& Hernquist 1994a; Taniguchi \& Wada 1996).
In particular, Bernlohr (1993) studied the starburst in 29 interacting
galaxies, mainly with unequal mass (fitting synthetic Starburst spectra to
the observed one). He found that the starburst in the minor galaxies
started earlier than in the major galaxies; and that the delay between the
burst of the components can be significantly larger. In NGC 5514 was found
that both extra--nuclear starburst started almost at the same time, or
even earlier in the major galaxy..

From detailed simulations of starbursts in major and minor mergers between
disk/bulges/halo galaxies, Mihos \& Hernquist (1994a,b, 1996) proposed that:
(1) in mergers of bulge dominant galaxies there is a significant increase
in the star formation rate, in relation to mergers of disk/halo systems;
(2) the central bulges act to stabilize the collision against radial
inflow and associated starbursts, until the galaxies merger. In other words,
they proposed--in the last point--that
in mergers between bulge dominant galaxies the
massive star formation occurred very late in the history of the interaction.
Our 2D studies of starburst and GW in IR mergers/QSOs suggest that this
proposition is probably true for the cases of  NGC 2623, Mrk 231, NGC 3256,
Arp 220, and others (Lipari et al. 2004a, 2000, 1994).
For NGC 5514 (where we verified that this system is a collision of
two bulge dominant galaxies) it was found two young extra--nuclear starbursts
processes with ages of $\sim$ 3--8 $\times$ 10$^6$ yr, then the starburst
activity was delayed until well after the tails were launched, as would be
expected for systems with substantial bulges. Probably the gas was
concentrated--in the central areas--earlier.

In conclusion,
in this work was presented the first, physical, morphological and
kinematics evidence that the two strong extra--nuclear emission areas in
NGC 5514 are associated to  starbursts + OF + LINERs, with a supergiant
galactic bubble.
In Section 4.4 other properties of these starburst structures will
be discussed.


\subsection{LINERs in NGC 5514 and IR mergers/QSOs}
\label{dis-LINER}

\subsubsection{LINER properties in the extra--nuclear starbursts and the
nuclei in NGC 5514}
\label{dis-LINER1}

The LINER properties found in the main nuclear and extra--nuclear regions
of NGC 5514 (obtained from the emission line maps/tables),
will be discussed in the next paragraphs:

\begin{enumerate}

\item
{\it Bubble Region}:
we found LINER characteristics in  all the supergiant galactic bubble
structures. In particular in all the knots and in the BBC and RBC
emission line systems (Tables 5 and 7).

Furthermore, in the bubble 100 per cent of the field shows very high
[N {\sc ii}]/H$\alpha$ and [S {\sc ii}]/H$\alpha$ ($>$ 0.8).
These ratios for the supergiant galactic bubble of
NGC\,5514 are located in the area of `fast shock velocities' (of
$\sim$300-500\,km\,s$^{-1}$), in the superior end part of the
[N {\sc ii}]$\lambda$6583/H$\alpha$ vs.
[S {\sc ii}]$\lambda$ 6717 + 31/H$\alpha$ diagram
(published by Dopita \& Sutherland 1995, their Fig. 3a).
These results could be associated mainly to the effects of shock
ionization when the out flowing gas collides with the ISM (Heckman 1980, 1996;
Heckman et al. 1987, 1990; Dopita 1994; Dopita \& Sutherland 1995;
Shull \& McKee 1979).

Similar results were obtained in the 2D studies of the bubble of NGC 3079
(Veilleux et al. 1994) and the OF nebula of NGC 2623 (L\'{\i}pari et al.
2004a). They found that
the ratios [N {\sc ii}]$\lambda$6583/H$\alpha$ and
[S {\sc ii}]$\lambda$ 6717 + 31/H$\alpha$ are $>$ 1, in  almost all the
bubble and the nebula.
They associated these results to the presence of large scale OF + shocks.

Furthermore, it is important to note that Lutz (2004, private communication,
1990) found that an interesting additional feature in NGC 5514: a strong radio continuum
emission (about an order of magnitude above the radio-FIR correlation), 
again mainly arising in the off-nuclear regions.
This feature has implications for the interpretation in the LINER+shocks
scenario. In particular, Lutz (1990) interpreted this as
a signature of direct particle acceleration in galaxy wide shocks
(favouring this explanation than other interpretations, like an unusual
AGN or the signature of sharply decaying star formation with far IR flux
already down, radio lagging behind).

\item
{\it Complex of H {\sc ii} Regions}:
at the border of this complex was also found  OF and
LINER properties (Sections 3.3, Tables 5 and 7).
Again these LINER characteristics are probably associated to giant shocks
in the compressed ISM (by the OF clouds).

For the emission peak in the main/central knot, of this H {\sc ii}
complex, the observed emission line ratios (Tables 5 and 7) are
consistent with transition objects between LINER and H {\sc ii} regions
(Ho 1996; Heckman 1996).
This transition characteristic, similar to that found for the second nucleus,
is probably associated with OF process, plus the superposition with H {\sc ii}
regions.

\item
{\it Main Nucleus}:
for this nuclear region, the optical emission line maps and spectra  show
values consistent with a weak LINER/shocks plus an old stellar population.
In addition, the surveys at X-ray wavelengths  found
no evidence of a strong emitter or an AGN.

However, using the NRAO VLA Radio Sky Survey (Condon et al. 1998) and the
IRAS Faint Source Catalogue (Moshier 1992) for the galaxies in the Uppsala
Catalogue, Condon, Catton \& Broderick (2002) derived for NGC 5514/UGC 9102 a
value of $q = log[F_{FIR}/S_{1.4 GHz}] =$ 1.96, where
$F_{FIR}$ and $S_{1.4 GHz}$ are the far IR flux and the radio flux density,
respectively. This low value of $q$ suggests that the dominant energy
source for the radio emission is from an AGN. They associated this value
of $q$ with the main spiral galaxy (Sa, NGC 5514--NED1) of the merger.

These results suggest that the nuclear source of bolometric energy
in the main nucleus, is probably dominated by an AGN plus shocks.
Thus, their LINER properties could be also associated mainly to
AGN + shocks processes.

\item
{\it Second Nucleus}:
it is important to analyse the nature of this nucleus where
properties of transition LINER/H {\sc ii} object were detected.

In particular,
it is interesting to discuss the result obtained using Tables~\ref{elr2dt}
and ~\ref{elr2df} for this second nucleus: if we study the properties of the
emission peak, in only one fibre (of diameter of $\sim$1$''$ $\sim$ 476 pc)
mainly LINER characteristics were observed.
However, if the circumnuclear region is included (in a diameter of $\sim$2$''$
around this nucleus) their spectrum depict transition LINER/H {\sc ii}
properties. Thus, for--at least--this case the transition characteristics
could be associated mainly to a spatial superposition of LINER plus
H {\sc ii} regions (see Barth \& Shields 2000; Heckman 1996; Ho 1996).

Finally, we note that the ELR of both nuclei (Tables 5 and 7)
are located in the
[O {\sc i}]$\lambda$6300/H$\alpha$ vs.
[S {\sc ii}]$\lambda$6717 + 31/H$\alpha$ diagram, in the same area as
to that found by Heckman et al. (1990) for the extra--nuclear region of a
sample of 14 IR galaxies with galactic winds (their Fig. 14).
Furthermore, the ELR of both nuclei of NGC 5514 are located (in this
plot) in the small area
of Herbig--Haro objects and supernova remnants: i.e., `the shocks region'

Therefore, with the data available for the second nucleus (of NGC 5514),
their LINER properties could be associated  mainly to star
formation and shocks processes.

\end{enumerate}

\subsubsection{LINER properties associated with starburst and galactic winds
in IR mergers/QSOs}
\label{dis-LINER1}

In the present programme of 2D spectroscopic study of IR mergers and IR QSOs
with galactic winds--and associated nebula or bubbles--we found a relatively
high number of objects that have LINER properties, associated mainly
with nuclear starburst and large scale shocks (see Section 4.5 and Lipari
et al. 2004a: their Table 1).
Furthermore, in NGC 5514  LINER properties in the  nuclear regions, and
very strong LINER activity in the extra--nuclear OF areas (Section 3)
were observed.

In addition, this new result for NGC 5514, confirms an interesting previous
result (obtained by Lipari et al. 2003, 2004a): mainly starbursts and LINERs
are the sources of ionization in `low velocity OF' IR mergers
(LVOF, $V_{\rm LVOF} <$ 700\,km\,s$^{-1}$; L\'{\i}pari et al.\ 2003, 2004a).
According to the results obtained from this and previous detailed 2D
spectroscopic studies of NGC\,5514, 2623, 3256 and similar IR mergers,
we already suggested that three processes could be the main sources
of ionization and energy in these LVOF IR systems:
(1) in nuclear and circumnuclear regions, the gas could be
photoionized by the diffuse radiation field of metal-rich dusty starbursts,
showing `starburst or LINER' spectra (Heckman 1996; Wang et al.\ 1997;
Ho 1996; Barth \& Shields 2000; Filippenko \&
Terlevich 1992; Shields 1992; Terlevich \& Melnick 1985);
(2) in very extended areas, the source
of ionization could be large scale shocks in clouds
accelerated outwards by starbursts + galactic winds, which produce LINER
properties, especially at the periphery of the OF areas (Heckman 1987, 1996;
Heckman et al.\ 1987, 1990; McCarthy et al.\ 1987; L\'{\i}pari et al. 2004,
in preparation); and
(3) in several regions, including the nuclei, these two previous processes
could be superposed.
These  sub-classes of LINER associated with nuclear
starburst + shocks, and with large scale shocks are clearly
different from the  `AGN-related LINER' sub-group (Heckman 1996;
Wang et al. 1997; Ho 1996).

The new results obtained from our programme could help us to understand
the nature of LINERs and the strong IR emission, in some LIRGs and ULIRGs.
Where LINERs are almost the dominant spectral types
($\sim$30--40 per cent; Sanders \& Mirabel 1996, their fig.\ 5).
Specifically, our results are consistent with those of Lutz et al.\ (1999)
and Veilleux et al.\ (1999): they suggested that the  nature of LINERs
in LIRGs and ULIRGs is associated mainly with starbursts and
shocks originated in galactic winds.


\subsection{Galactic winds in NGC\,5514.}
\label{dis-gw55}

\subsubsection{The supergiant galactic bubble}
\label{dis-gw1}

Fig.~\ref{bubblec} depicts a schematic diagram of the supergiant galactic
bubble. This plot shows the emission line morphology plus the main ejections
detected in NGC 5514 (Sections 3.5 and 3.3).
We note that these results are in agreement
with the general scenario  proposed for the OF process by  hydrodynamics
GW models (e.g., Suchkov et al. 1994; Strickland \& Stevens 2000). In particular,
the result obtained in Section 3 agree with: the `pre--blowout' or rupture
phase of the bubble (see the summary of GW phases, presented in Section 1.2).

Following the study of the superbubble in NGC 3079 (Veilleux et al. 1994)
the dynamical timescale of the bubble in NGC 5514 was derived, using the
relation:\\

t$_{dyn} =$ 1.0 $\times 10^6$ R$_{bubble, kpc}$ V$^{-1}_{bubble, 1000}$ yr

where  R$_{bubble, kpc}$ and V$^{-1}_{bubble, 1000}$ are the linear dimension
of the bubble and the velocity of the entrained material (in units of kpc and
1000 km s$^{-1}$, respectively). A value of  t$_{dyn} \sim$ 8.5 $\times$
10$^6$ yr was obtained.

A second age can be estimated if the bubble has not reached the blowout phase
(as the case of NGC 5514), and can be approximately like an adiabatic structure
with a radiative shell that expands through an uniform medium
(Koo \& McKee 1992).
Then by using the relations for the radius and expansion velocity of the shell
(from Castor, McCray \& Weaver 1975; Veilleux et al. 1994) they derived:\\

t$_{bubble} =$ 6.0 $\times 10^5$ R$_{bubble, kpc}$ V$^{-1}_{bubble, 1000}$ yr

a value t$_{bubble} \sim$ 5.2 $\times$ 10$^6$ yr was found, very similar to
the dynamical age.
This derived age for the bubble allows the formation of the SSCs in
the external/compact knots, with ages less than 6 $\times$ 10$^6$ yr.

In Section 3.5 the following OF velocities were measured close to the
centre of this supergiant galactic bubble:
$V_{OF blue}=(-320\pm20)$\,km\,s$^{-1}$ and
$V_{OF red}=(+265\pm25)$\,km\,s$^{-1}$.
These velocities measured for the strong blue and red bubble components are
consistent with expansion of the shell, and together with the  clear
ejections suggest that the OF will reach the blowout phase.
The velocities measured in the ejection areas and even for the expansion of
the shell are higher than or similar to the escape velocity in extra--nuclear
regions (e.g., the escape velocity for the solar neighbourhood  is
$V_{escape} =$ 350 km s$^{-1}$; Binney \& Tremaine 1987).

From the 2D spectra, for the blue and red bubble emission line components a
H$\alpha$ flux of  1.1 $\times$ 10$^{-13}$ erg cm$^{-2}$ s$^{-1}$ was derived.
Assuming a mean value $E(B-V)= 0.44$ (from Tables 4 and 6) and using
 the relations given by Mendes de Olivera et al.\ (1998) and
Colina et al.\ (1991) we found a value for the ionized-gas mass, in the
OF, of 4.2 $\times$ 10$^{6}$ $M_{\odot}$.
In IR galaxies, different studies have found that the range of ionized-gas
mass ejected by the OF is $\sim$10$^{6}$--10$^{8}$ $M_{\odot}$ (Lipari et al.
2004a; Colina et al.\ 1991; Veilleux et al.\ 1994).
Using the value of ionized-gas mass, we derived the kinetic energy of the
OF, E$_{\rm KIN-OF} \sim$  0.5 $\times$ M$_{\rm OF-IG}$ $\times$
$\langle V_{\rm OF}\rangle^{2}$ = 4.0 $\times$ 10$^{54}$ erg.
This result is within the range obtained for the OF in luminous IR
mergers, M82, and NGC\,3079  (Lipari et al. 2004a).

Finally, it is important to remark that from our 2D INTEGRAL study
(Sections 3.3 and 3.5)
direct evidence of a very high number of massive WR stars were found,
in the OF knots of the NGC 5514 broken bubble.
These WR stars are  progenitors of core--collapse super or even
hypernova.  Thus, as it is suggested by theoretical works, SNe from massive
progenitors are probably the objects that generated the rupture phase of
the bubble, in the OF knots  (for references see Section 4.5.2).

\subsubsection{The outflow in the H {\sc ii} region complex}
\label{dis-gw2}

The general morphology and the spectra of this H {\sc ii} region complex,
clearly suggest that the star formation and the OF process are
very young: with an age less than 3--4 $\times$ 10$^6$ yr (i.e., in
the first phase of the starburst process, which is dominated by hot main
sequence stars with H{\sc ii} region spectra; Terlevich et al. 1993).

In the external areas of the complex of H {\sc ii} regions we derived for
the OF a H$\alpha$ flux of  2.0 $\times$ 10$^{-14}$ erg cm$^{-2}$ s$^{-1}$
(which is $\sim$25 per cent of the total flux in this area).
Assuming a mean value $E(B-V)= 0.48$ (from Tables 4 and 6) and using
 also the relations given by Mendes de Olivera et al.\ (1998) and
Colina et al.\ (1991)  a value for the ionized-gas mass, in the
OF, of 1.0 $\times$ 10$^{6}$ $M_{\odot}$ was found.
Using this value of ionized-gas mass, we derived for the kinetic energy of
the OF, E$_{\rm KIN-OF} \sim$  0.5 $\times$ M$_{\rm OF-IG}$ $\times$
$\langle V_{\rm OF}\rangle^{2}$ = 0.8 $\times$ 10$^{54}$ erg.

In Section 3.5  for the weak OF components (detected at the border of this
complex of H {\sc ii} regions), the following OF velocities were measured:
$V_{OF blue}=(-240\pm25)$\,km\,s$^{-1}$ and
$V_{OF red}=(+100\pm30)$\,km\,s$^{-1}$.
We note that this young OF process depicts some properties similar to those
observed in the OF process associated with the supergiant galactic bubble
(e.g., a major axis of $\sim$5 kpc, LINER+shocks
properties at the border/shell, etc).
Thus, this young OF could generate a new giant galactic bubble, in NGC 5514.
However, the H$\alpha$ luminosity in the main knot (of this complex) is
less--in almost one order of magnitude--than the luminosity in the
knot 1 of the bubble.

One of the more interesting  feature in this area
is the circular structure observed in [S {\sc ii}]/H$\alpha$, and the
velocity field map (see Sections 3.5 and 3.6).
This structure  is probably associated with the interaction of OFs processes
detected in this merger. Probably this OF--that we are looking blushifted and
face on--could be generated in  the far side of the OF process detected in
the west area.
In other words, the west bubble could be the nearest part of a bipolar OF.
This type of structure, with double bubble, was already detected in NGC 3079,
30 Dor,  NGC 2782, NGC 2623, etc (Ford et al. 1986; Meaburn 1980; Redman
et al. 2003; Jogee et al. 1998; Lipari et al. 2004a).

Finally, it is important to remark another interesting result found
in Section 3 (for NGC 5514) and in our programme (for NGC 3256 and
NGC 2623):
{\it `the [S {\sc ii}] and  [S {\sc ii}]/H$\alpha$ are probably the
best tracers of GW and OF processes'}.  We recall that
a good evidence of shock process and SN events (in
ionized nebulae) is the intensity enhancement of low excitation lines,
in particular [S {\sc ii}] and [O {\sc i}]. In fact, the
[S {\sc ii}]/H$\alpha$ and [O {\sc i}]/H$\alpha$ ratios have been
used as the best tracer of shock ionization mechanisms (Heckman 1980;
Canto 1984; Binette 1985; Heckman et al. 1990; Masegosa, Moles \& del
Olmo 1991; and others).
Specifically, in Herbig--Haro objects the OFs and shocks processes  were
detected mainly/first in [S {\sc ii}] emission and [S {\sc ii}]/H$\alpha$
ratio (Canto 1984).
In IR mergers and IR QSOs, different works already proposed that giant
galactic shocks are associated with extreme starburst + galactic winds,
induced by the merger events
(Joseph \& Wright 1985; Riecke et al. 1985; Heckman et al. 1987, 1990;
Armus et al. 1988; Lipari et al. 1993, 1994, 2000, 2003, 2004a; Lutz et
al. 1999; Veilleux et al. 1999).


\subsection{Galactic winds in IR mergers and IR QSOs}
\label{dis-gwir}

It was found in this work strong evidence of two extra--nuclear galactic
winds in the IR merger NGC 5514, plus an AGN in the main nucleus.
Thus this object is a new example of IR merger with composite activity:
extreme starburst plus AGN. It is important to remark, that
two of the main goals of our programme are to study this type of composite IR
systems, and the possible links--or evolutionary paths--among IR mergers,
extreme starburst + GW, and AGNs/QSOs.
In this section these two main points are discussed. In addition, 
interesting results obtained from the last update of the database
of our programme are presented.

\subsubsection{Galactic winds in IR mergers}
\label{dis-gw3}

The low velocity OF (LVOF, $V_{\rm EVOF} <$ 700 km s$^{-1}$) observed
in the extra--nuclear starbursts of NGC\,5514 is
consistent with the results found in IR mergers with dominant massive
starbursts processes--e.g., in NGC\,2623, NGC\,3256, NGC\,4039/38,
and others--where  mainly LVOF were detected (L\'{\i}pari et al.\ 2000,
2003, 2004a).
This fact also suggests the importance of studying, as a group, the
general properties of these IR systems.

Lipari et al. (2004a) found that a high proportion of IR mergers develop
galactic winds: at least $\sim$75\%. This interesting result  was found
comparing their data base of OF in IR mergers/QSOs with
two samples of nearby IR merger.
Thus, this result strongly suggests---or confirms---that: (a) GWs are
`frequent events' in IR mergers; (b) extreme starbursts + GW and extreme
IR emission could be simultaneous processes induced by merger events.

This last conclusion is also supported by a clear trend found in the plot
OF velocity vs.\ $\log L_{\rm IR}$ (for  28 IR mergers/QSOs with
confirmed GW), in the sense that
extreme OF velocities are detected only in extreme IR emitters (ULIRGs).
The proposed explanation for this observed trend is that high values of OF
velocity and $L_{\rm IR}$  are both associated mainly with the same process:
`starburst + QSO' events, probably induced by mergers (Lipari et al. 2004a).

\subsubsection{Galactic winds in IR QSOs}
\label{dis-gw4}

In previous works we stressed the fact that it is important to study
if `some' IR mergers (like NGC 5514) with starbursts and low velocity
OFs could evolve into IR QSOs with composite nuclear
nature and extreme velocity OF, or to bulge/elliptical galaxies.
This possible evolutionary path is supported by several facts,
in particular the presence of a correlation between the mass of galactic
bulge and  the mass of supermasive black hole candidate (or supermasive
dark object;  Kormendy \& Richstone 1995;
Laor 1998; Merrit \& Ferrarese 2001), strongly suggests that the
formation and evolution of  bulges, supermasive black hole, and QSOs 
are physically related to each other (Kawakatu et al. 2003).

Recently, for IR QSOs we found:
(i) extreme velocity OF (EVOF, $V_{\rm EVOF} >$ 700 km s$^{-1}$) objects
with a composite nuclear source: starbursts + AGNs/QSOs; for e.g.
in Mrk\,231, IRAS 19254-7245, 01003-2230, 13218+0552, 11119+3257, 14394+5332,
and others (L\'{\i}pari et al.\ 2003);
(ii) high resolution {\itshape HST\/} WFPC2 images of IR  + BAL +
Fe\ {\sc ii} QSOs show in practically all of these
objects arcs or shell features probably associated with galactic
winds or merger processes (L\'{\i}pari et al.\ 2003).

Our evolutionary model for young/composite IR QSOs suggests that part of the
BAL systems and extreme Fe\ {\sc ii} emission could be linked to violent
supermassive starbursts (probably generated by mergers), which can lead
to a large scale expanding shell, often obscured by dust.

Therefore it is important to analyse different types of evidence for OF in
IR QSOs. Recently, Zheng et al (2002) reported a very interesting new
result: they measured the offset between the broad and narrow H$\beta$ and
[O {\sc iii}]$\lambda$5007 emission line components in a sample of 25 IR QSOs;
and they found clear offset in $\sim$76 per cent of their sample. They
associated this offset to OF processes.
We note, that previously Taniguchi et al (1994) performed a similar study
for the broad and narrow Pa$\alpha$ emission line in the IR QSO
IRAS 07598+6508; and they already suggested that this type of
offset could be associated to OF events.
We have verified that for Mrk 231 the results using the offset method
(by Zheng et al. 2002) gave the same value of OF: --1000 km s$^{-1}$,
to that already obtained from the detection of two emission line
systems in [O {\sc ii}]$\lambda$3727 (by Lipari et al. 1994).
Therefore we followed this offset method (OSM) to study our OF IR QSO
candidates.

We recall that this OSM requires two important steps: first, for objects
with Fe {\sc ii} emission it is required the subtraction of a Fe {\sc ii}
template (using for example the Boroson \& Green 1992 procedure).
Then, it is required a careful deblending of the broad and narrow emission
line components (using for example the detailed SPECFIT software package;
see for detail Zheng et al 2002; Lipari et al. 2003, Lipari 2004, and Section 2).

Table~\ref{gwmergersqso} presents the updated values of our data base of
OF in  IR QSOs. Including 12 new values of OF obtained by using this offset
method and applied to our OF IR QSO candidates (from Lipari et al 2004a:
Table 1).
In Table~\ref{gwmergersqso} was also included the offset--OF values for
15 IR QSOs reported by Zheng et al. (2002). We selected objects with
$\vert$offset--OF$\vert$  $>$ 200 km s$^{-1}$.
In addition, it was included a new column, with the values of the emission
line ratio of Fe {\sc ii}$\lambda$4570/H$\beta$ (RF).
We recall that this Table presents the main properties of IR mergers/QSOs
with galactic wind or candidates (including the results obtained for
NGC\,5514).

From this Table~\ref{gwmergersqso} we  remark 3 interesting results:
(1) All the measured offset -from objects with and without Fe {\sc ii}
emission- show that the H$\beta$ broad emission line are blue shifted,
in relation to the H$\beta$ narrow component.
(2) Plotting  Fe {\sc ii}$\lambda$4570/H$\beta$ vs. offset--OF
(Fig.~\ref{feof}), including our
8 new objects with measured values of offset and Fe {\sc ii} emission, we
confirm the trend found in this plot (already published by
Zheng et al. 2002), in the sense that extreme OF velocities are detected
only in strong and extreme Fe {\sc ii} emitters. Lipari et al. (1993)
already defined strong and extreme Fe {\sc ii} emitters as objects with the
ratio Fe {\sc ii}$\lambda$4570/H$\beta$ $>$ 1 and 2, respectively.
(3) It is clear in this Table, the fact, already noted by
Taniguchi et al. (1994), that in cold ULIRGs there is an absence of
Fe {\sc ii} emission. Thus it is interesting to analyse if this fact
means the absence of type II SN or  the starburst.
In relation to this last point, Terlevich, Lipari \& Sodre (2004, in
preparation) note that ULIRGs are very dusty objects with mainly
LINER and Syfert 2 spectra. However, when it is observed the broad line
region (BLR) in ULIRGs, it is frequently observed also the Fe {\sc ii}
emission.

The detection in all our OF IR QSO candidates that the H$\beta$ broad line
component is blushifted in relation to the narrow one, it is clearly consistent
with the result obtained from the study of strong Fe {\sc ii} + BAL
emitters, by Boroson \& Meyer (1992). They proposed that blueshifted
offset/asymmetry detected in the H$\alpha$ broad  components--of IR QSOs
with strong Fe {\sc ii} + BAL systems--is probably due to
the emission of the out flowing material, associated with the BAL process.
Furthermore, our measurements are also consistent with the results
of the study of multiple emission line
components, in some starburst nucleus of galaxies (Taniguchi  1987).
Thus, a possible explanation for
at least part of these blushifted broad line systems in IR QSOs
is the high speed OF/GW generated in a burst of SN event near the
nuclear region (Heckman et al. 1990; Terlevich et al. 1992; Perry \&
Dyson 1992; Lipari et al. 2003, 1994, 2003, 2004a; Taniguchi et al. 1994;
Scoville \& Norman 1996; Lawrence et al. 1997; Collin \& Joly 2000; and
others)

In relation to the clear trend found (or confirmed) in this paper,
between the extreme offset--OF and Fe {\sc ii} emission, it is
important to recall:

\begin{enumerate}

\item
We suggested that IR QSOs could be {\it young IR
active galaxies at the end phase of an extreme starburst: i.e.\ composite
and transition IR QSOs}.
At the final stage of an `extreme starburst', i.e. type II SN phase
([8--60]\ $\times10^{6}$\,yr from the initial burst; Terlevich et al.\ 1992,
1993) powerful galactic winds, giant galactic arcs, BAL systems and extreme
Fe\,{\sc ii} emission can appear (see for references L\'{\i}pari et al.\
2004a).

\item
In the shell or arc model proposed for  Fe {\sc ii} + BAL + IR QSOs, the high
fraction of IR QSOs showing properties of low ionization BAL + Fe {\sc ii}
systems (Low et al. 1989; Boroson \& Meyers 1992; L\'{\i}pari 1994) could be
explained by the high fraction of arcs, shells, and giant SN rings
present in these IR systems (probably originated in the starburst type II
SN phase).

\item
Recently, several studies have confirmed the `composite nuclear nature' and
the presence of merger features in several IR QSOs.
In particular, using Keck spectroscopy,
Canalizo \& Stockton (2001) have studied mainly IR
QSOs  defined as {\it transition/composite} objects
between ULIRGs and standard QSOs (L\'{\i}pari 1994: fig.\ 5).
They found clear spectroscopic evidence of strong {\it young stellar
populations}, in the host galaxies of these systems.
They also detected that, out of the nine transitional or
composite IR QSOs studied, eight are major mergers.

\item
Type II SNe are highly concentrated in space and time and arise from
massive stars (m $\geq$ 8 M$_{\odot}$) in young stellar clusters and
association (of ten and hundreds of massive stars; Heiles 1987).
Thus, the presence of association of type II SN as a result of the evolution
of massive stars should be expected (Norman \& Ikeuchi
1989; Hillebrand 1986; Woosley \& Weaver 1986). In particular, WR massive
stars are progenitors of core--collapse SNe.

Theoretical and observational works suggest that Type II SNe are the main
galactic objects capable of generating the blowout
phase of the galactic winds (see for references Norman \& Ikeuchi 1989).

\end{enumerate}

It is important to remark that the presence of {\it strong} Fe {\sc ii}
emission requires `very specific physical conditions', in the clouds of
the broad and narrow line regions.
In particular, Joly (1987) found--for the BLR--that high Fe {\sc ii}/H$\beta$
ratio could be explained mainly with purely collisional models showing
low temperature (T $<$ 8000 K), very high density (N$_e$ $>$ 10$^{11}$
cm$^{-3}$), and high column density (N$_H$ $>$ 10$^{22}$ cm$^{-2}$).
For the narrow line region,
Verner et al. (2002, 2000) and Veron, Joly \& Veron (2004) have shown
that the presence of permitted (and forbidden) Fe {\sc ii} requires high or
intermediate density (10$^{6}$ $<$ N$_e$ $<$ 10$^{8}$ cm$^{-3}$) and
intermediate ionization parameter (10$^{-6}$ $<$ U $<$ 10$^{-5}$).
Thus, these results suggest that the `strong' Fe {\sc ii} emission could be
detected in a relatively low percentage of objects or cases.
However, it is very interesting to note that $\eta$-Carinae, the nearest
object with a strong OF, has also strong Fe {\sc ii} emission (see Gaviola
1953; Verner et al. 2002; Veron et al. 2004).
More specifically, the core (0\farcs3) of $\eta$-Carinae was resolved by
speckle interferometry into A, B, C and D components (Weigelt \& Ebersverger
1986). {\it HST} observations found that the brightest component A is the
central massive star+wind and the other objects (B, C, and D) are slow
moving ejecta (Davidson et al. 1995, 1997; Gull, Ishibashi \& Davidson 1999).
The {\it HST/STIS}--spectrum of the source A  is rich in broad
permitted Fe {\sc ii} lines, with many P Cygni absorption components
(Hiller et al. 2001). Narrow permitted and
forbidden Fe {\sc ii} dominated in the B-D spectrum.
The gaseous B, C, and D blobs have
low temperature T$_e$ $\sim$ 6500 K  and relatively high density
N$_e$ $\sim$ 10$^{6}$ cm$^{-3}$, in agreement with the results of the
Fe {\sc ii} models.

In conclusion, the presence of starbursts + GWs/OFs + type II SNe in IR QSOs
could be a possible explanation for `part' of the trend found in the plot
Fe {\sc ii}/H$\beta$ vs. offset--OF (Fig.~\ref{feof}).
More specifically, this trend (in the sense that in IR QSOs extreme
offset--OF velocities are detected only in strong and extreme Fe {\sc ii}
emitters), could be explained by the fact that in IR QSOs a combination
of the following 3 processes are working:
(1) large scale giant ionizing OF--shocks;
(2) association of massive stars that probably end their evolution as
type II SNe, and this process can enrich with Fe the ISM and the BLR;
(3) obscured QSOs that gradually ionize part of the BLR (as suggested by the
standard models of photoionization, by a supermasive black hole).
Previously, we found  clear evidence that extreme OF events are associated
with these 3 `main processes' (Lipari et al. 1994, 2003, 2004a).

Finally, we note that the results of the present 2D
spectroscopic study of {\it `IR Mergers and IR QSOs with Galactic Winds'}
will be used in a second part of this programme, which is devoted
to the study of galactic wind in Ly$\alpha$ emitters  at high redshift.
This study is based on a very deep imaging and spectroscopic survey for
Ly$\alpha$ emitters at z=5.7 using the Suprime-Cam and GMOS
on the Subaru and Gemini telescopes (Taniguchi et al. 2003; Ajiki et al. 2002,
2003; Lipari et al. 2004b).
This survey gave us a well-defined sample of 20 LAE candidates
at z = 5.7 and two of them were  confirmed as LAEs at z = 5.7
(Ajiki et al. 2002; Taniguchi et al. 2003).
Furthermore, Ajiki et al. (2002) found in the optical Keck II spectrum
of J1044--0130 (at z = 5.687) the typical line profile  of galactic wind;
thus this object is probably the more distant GW observed to date.
Therefore, it will be interesting to compare the results of the present 2D
spectroscopic/imaging study of {\it `galactic winds'} at low redshift
(for NGC 2623, NGC 5514, NGC 3256, Mrk 231, Arp 220, and others) with those
obtained from the imaging and 1D spectroscopic analysis of galactic winds
at high redshift (when the galaxies and QSO formed).

In part, this type of study and comparison were already started and performed.
For example, Hibbard \& Vacca (1997) used rest--frame  B and V images of
nearby interacting/mergers and starburst galaxies (e.g., NGC 1614,
1741, 3690, He 2-10) to simulate {\it Hubble Deep Field} observations
in F606W and F814W filters of starburst galaxies in the redshift range
z $\sim$ 0.5--2.5.

It is important to recall that the same problem of lack of
spatial resolution in the analysis of galaxies/QSOs at high redshift,  is
present in the studies of the BLR and supermasive black hole in galaxies/QSOs
at low redshift. Therefore, it is interesting to answer a simple
question: how will NGC 5514 look at redshift z $>$ 3 ?
(with 2 extra--nuclear region with strong LINER activity).
Probably, this merger could be detected mainly as a point source with
multiple or broad emission line systems, and properties of a transition
LINER/H {\sc ii} object. These characteristics are attributed  frequently
to  the nuclear region and  the BLR.

\clearpage

\section{Summary and Conclusions}

In this paper a study of morphology, kinematics and ionization structure
of the IR merger NGC\,5514 is presented. The study is based
on 2D INTEGRAL spectroscopy (obtained on 4.2 m WHT, at La Palma
Observatory) and long--slit spectroscopy  plus broad band images
(obtained on CASLEO and CTIO 2.15 and 1.0 m telescopes, respectively).
The main results and conclusions may be summarized as follows:

\begin{enumerate}

\item
Clear evidence of two extra--nuclear starburst regions with young outflows
 and LINER activity are reported.
One of these OFs generated a {\it supergiant galactic bubble detected just
in the rupture and pre--blowout phase} and the other one
is associated with an extended complex of giant H {\sc ii} regions.

\item
For the galactic bubble was found:
(1) the [S {\sc ii}], H$\alpha$, [N {\sc ii}], [O {\sc i}] and [O {\sc iii}]
emission line maps show clearly a supergiant bubble with ellipsoidal shape,
with major and minor axes of  $\sim$6.5 kpc (13\farcs6;
at PA $=$ 120$^{\circ}\pm10^{\circ}$) and $\sim$4.5 kpc (9\farcs6);
(2) the centre is located at 8\farcs5 $\sim$4.1 kpc, to the
west from the main nucleus;
(3) there are 4 main knots, one very strong/extended and three more compact
at the border;
(4) the WHT spectra show two strong components, (blue and
red emission line systems), probably associated with the emission from the
near and far side of the external shell of the bubble;
(5) these two components depict LINER properties, probably associated with
large scale OF + shocks;
(6)  at the east border, the kinematics and the [S {\sc ii}] emission line
maps show a strong and extended ejection of ionized gas, aligned
with the PA of the major axis;
(7) another 3 ejections from the bubble were found, two of them are
perpendicular to the extended one.
These results strongly suggest that the GW/OF process is in
the rupture of the bubble and in the beginning of the blowout phase.

\item
In the knots (of the bubble) 1, 2a, 2b, 3a and 3b were detected Wolf-Rayet
features/bump at $\lambda$4560 \AA. In addition, only in these knots (where
we detected the WR bump) were also found the [N {\sc i}]$\lambda$5199
emission line. A similar behaviour, i.e. strong WR and [N {\sc i}] emission,
was observed in the spectra of the proto--type of WR galaxy, NGC 6754.

\item
For the extended complex of giant H {\sc ii} regions was detected:
(1) the H$\alpha$, [N {\sc ii}]$\lambda$6583 and
[S {\sc ii}]$\lambda\lambda$6717 + 31 emission line maps show
a compact area of strong emission line (with a peak of emission located at
1\farcs7 $\sim$810 pc, to the east of the second nucleus), and a faint
extended emission with elongated shape,
major and minor axes of  $\sim$5.1 kpc (10\farcs8;
at PA $\sim$20$^{\circ}$) and $\sim$2.9 kpc (6\farcs0);
(2) inside of this complex, the spectra show H {\sc ii} region and H {\sc ii}/LINER
characteristics; (3) however at the border of this extended area the spectra
have outflow components and LINER properties (similar  to those observed
in the OF of the galactic bubble).

\item
INTEGRAL 2D [N {\sc ii}], H$\alpha$, [S {\sc ii}] and [O {\sc iii}]
velocity fields (VFs) for the main body of the merger  are presented
(mosaics covering the area
$\sim$30$''\times$20$''$; 14.3 kpc$\times$9.5 kpc).
These VF maps show results consistent with an expansion
of the bubble, plus several strong and weak ejections.

\item
The U, B, V, J, H and K$_S$ images show a pre--merger morphology, from which
faint filaments ($\sim$2.8 kpc) of emission emerge from the bubble
position.

\item
The ionization structure and the physical conditions were analysed,
using 2D emission line ratios and width maps:
[S {\sc ii}]/H$\alpha$, [N {\sc ii}]/H$\alpha$,
[O {\sc i}]/H$\alpha$, [O {\sc iii}]/H$\beta$ and FWHM--[N {\sc ii}].
In the region of the bubble, 100 per cent of the field shows very high
[N {\sc ii}]/H$\alpha$ and [S {\sc ii}]/H$\alpha$ ($>$ 0.8).

\end{enumerate}

Thus the properties observed in these two extra--nuclear regions of
NGC\,5514 are consistent with two `early' galactic wind processes, which are
powered mainly by two starbursts (probably generated in the ongoing merger
process).
These new results for NGC\,5514 confirm our previous proposition that
extreme nuclear and extra--nuclear galactic winds processes are important
events in the evolution of IR mergers and IR QSOs.

Finally, the updated values of our data base of OF in IR QSOs/mergers
were presented. Including 12 new values of OF obtained by using an offset
method (for the narrow and broad H$\beta$ emission line), applied to our
OF IR QSO candidates.
In addition, in this data base were  included the offset--OF values for
15 IR QSOs reported by Zheng et al. (2002).
All the measured offset show that the broad emission line are blue shifted,
in relation to the narrow component of the same emission line.
Plotting  Fe {\sc ii}$\lambda$4570/H$\beta$ vs. offset--OF we confirm the
trend previously found in this plot (Zheng et al. 2002),  
in the sense that extreme OF velocities are detected
only in strong and extreme Fe {\sc ii} emitters.
This trend could be explained by the fact that in IR QSOs are probably
working a combination of the following 3 processes:
(1) large scale giant ionizing OF--shocks;
(2) association of massive star that probably end their evolution as
type II SNe, and this process can enrich with Fe the ISM and the BLR;
(3) obscured QSOs that gradually ionize part of the BLR (as suggested by the
standard models of photoionization, by a supermasive black hole).
Previously, we found  clear evidence that extreme OF events are associated
with these 3 `main processes'.

\section*{Acknowledgments}

The authors thank M. Ajiki, J. J. Claria, J. C. Forte, L. Hernquist, J.
Laborde, B. Madore, T. Mahoney, X. Y. Xia, W. Zheng, for useful discussions
and assistance.
We wish also to thank T. Boroson, R. Green, D. Kim, D. Macchetto, D. Sanders,
S. Veilleux, M. Veron, and B. Wilkes for their spectra of IR QSOs,
kindly made available to us.
This paper is based on observations obtained at William Herschel Telescope
(Canary Island of La Palma, Spain), CASLEO (San Juan, Argentina) and CTIO
(La Serena, Chile) with the 4.2, 2.15 and 1.0 m telescopes.
The 4.2\,m WHT is operated by the Isaac Newton Group
at the Observatorio de Roque de los Muchachos of the Instituto de Astrofisica
de Canarias (IAC). The authors thank all the staff at the Observatories and
the Instituto for their kind support (S. L. \& R. D. also thank the IAC for
financial support).
This research was made using the
NASA Extragalactic Database NED, which is operated by the Jet
Propulsion Laboratory, California Institute of Technology, under
contract with NASA. This paper was supported in part by Grants from
CONICET, SeCyT-UNC, and Fundaci\'on Antorchas (Argentina).
Finally, we wish to thank the referee for constructive and
valuable comments.

\clearpage

\vspace{100mm}

\begin{table}
\footnotesize \caption{Journal of observations}
\label{observations}
\begin{tabular}{ccllcl}
\hline
\hline
Date           & Telescope/ & Type & Spectral Region &Expos. Time   & Comments \\
               & Instrument &      &               & seconds      & \\
\hline
               &             &      &                             &              &  \\
               &             &      &                             &              &  \\

1989 Jun 29   &2.15m CASLEO/Z-machine&aperture Sp&$\lambda \lambda$4700-7200 \AA&1500$\times$3&
aperture: 3$''\times$6$''$ \\
               &             &      &                             &              &  \\
               &             &      &                             &              &  \\

1990 March 9   &1.0m CTIO/2D FRUTTI&long--slit Sp&$\lambda \lambda$3600-7000 \AA&1800$\times$3&
PA 90$^{\circ}$ slit width= 1\farcs5 \\
               &             &      &                             &              &  \\
               &             &      &                             &              &  \\

1993 Jul 13   &2.15m CASLEO/UCS &long--slit Sp&$\lambda \lambda$6500-7000 \AA&1800$\times$2&
PA 90$^{\circ}$ slit width= 2\farcs0 \\
               &             &      &                             &              &  \\
               &             &      &                             &              &  \\

1997 Mar 12    &2.15m CASLEO/CCD Cam & Images&B                   & 240$\times$2
& seeing $\sim$2\farcs0 (FWHM) \\
"              &"            & Images&V                           & 300$\times$2 & "\\
"              &"            & Images&I                           & 300$\times$2 & "\\
1997 Mar 13    &2.15m CASLEO/UCS&long--slit Sp&$\lambda \lambda$4000-7500 \AA&1800$\times$2&
PA 90$^{\circ}$ slit width= 2\farcs5 \\

               &             &      &                             &              &  \\
               &             &      &                             &              &  \\
2000 May 11    &2.15m CASLEO/CCD Cam &Images&U                    & 180$\times$2
& seeing $\sim$1\farcs7 (FWHM) \\
"              &"            &Images&I                            & 180$\times$2 & "\\

2000 May 25    &2.15m CASLEO/UCS&long--slit Sp&$\lambda \lambda$6100-7200 \AA&2700$\times$2&
PA 90$^{\circ}$ slit width = 2\farcs0 \\
"              &"        &long--slit Sp&$\lambda \lambda$4700-5800 \AA &2700$\times$2&
" \\
2000 May 26    &"        &long--slit Sp&$\lambda \lambda$3800-7300 \AA&1800$\times$2&
PA 90$^{\circ}$ slit width = 1\farcs5 \\

               &             &      &                             &              &  \\
               &             &      &                             &              &  \\

2001 Apr 11    &4.2m WHT/Integral&2D Spect. &$\lambda \lambda$6000-7400 \AA &1800$\times$3&
 Position 1 (centred in the bubble)\\
"              &"        &2D Spect. &$\lambda \lambda$6000-7400 \AA &1800$\times$3&
 Position 2 (centred in the nuclei area)\\
2001 Apr 12    &"        &2D Spect. &$\lambda \lambda$6000-7400 \AA &1500$\times$3&
 Position 3 (centred in the H {\sc ii} complex)\\
"              &"        &2D Spect. &$\lambda \lambda$4500-5900 \AA &1800$\times$3&
 Position 3 \\
"              &"        &2D Spect. &$\lambda \lambda$4500-5900 \AA &1800$\times$3&
 Position 1 \\
               &             &      &                           &                 &  \\
               &             &      &                           &                 &  \\

\hline

\end{tabular}
\end{table}

\clearpage

\begin{table}
\footnotesize \caption{Fluxes of emission lines of NGC\,5514, from long--slit
spectroscopy (medium and high resolution; from CASLEO)}
\label{flux1d}
\begin{tabular}{lcccccc}
\hline
\hline

Lines &         &Fluxes$^{a}$ (M.Res.$^{b}$)&    &  &Fluxes$^{a}$ (H.Res.$^{c}$) & \\
EqW, FWHM, Lum&Main Nuc.&West Area&Sec.Nuc.+H {\sc ii} Reg.&Main Nuc.&West Area&Sec.Nuc.+H {\sc ii} Reg.\\

\hline
                             &       &     &      &      &     &      \\

H$\beta\lambda4861$          & absor.& 1.8 & 1.4  & (0.4)& 1.8 & 1.0   \\

[O {\sc iii}]$\lambda5007$   & 1.4 & 2.9  & 1.5   &  1.1 & 3.0 & 1.2   \\

[O {\sc i}]$\lambda6300$     &  1.2 & 2.0 & 1.4   &  0.8 & 2.4 & 1.1    \\

H$\alpha\lambda6563$         &  1.6 & 5.2 & 6.3   &  1.4 & 5.2 & 4.8    \\

[N {\sc ii}]$\lambda6583$    &  4.3 & 8.7 & 2.0   &  2.4 & 9.1 & 2.8    \\

[S {\sc ii}]$\lambda6717$    &  1.2 & 3.6 & 2.5   &  0.5 & 4.1 & 1.5    \\

[S {\sc ii}]$\lambda6731$    &  1.0 & 2.6 & 2.4   &  0.4 & 3.3 & 1.3    \\

                             &      &     &       &      &     &         \\

H$\alpha$/H$\beta$           & ---  & 2.9 & 4.5   & (3.5)& 2.9 & 4.8     \\
E(B-V)$_{I}$                 & ---  & 0.0 & 0.4   & (0.2)& 0.0 & 0.5     \\
                             &      &     &       &      &     &         \\
EqW H$\beta$ (\AA)           & ---  & 32.0& 23.0  & (2.9)& 33.0& 29.0    \\
EqW [O {\sc iii}] (\AA)      & 4.9  & 57.0& 24.9  &  6.5 & 51.0& 40.0    \\
EqW H$\alpha$ (\AA)          & 6.4  & 74.4& 95.0  &  5.8 & 50.0& 51.0    \\
EqW [N {\sc ii}] (\AA)       &13.3  &126.0& 45.0  & 10.5 & 93.0& 25.0    \\
                             &      &     &       &      &     &         \\
FWHM H$\alpha$ (km/s)        & 365  & 550 & 230   & 340  & 480 & 190     \\
FWHM [N {\sc ii}] (km/s)     & 510  & 595 & 300   & 470  & 515 & 250     \\
                             &      &     &       &      &     &         \\
Lum H$\alpha$$^{d}$          & 1.8  & 6.0 & 7.2   & 1.6  & 5.7 & 5.5     \\
Lum [N {\sc ii}]$\lambda6583$& 4.9  &10.0 & 2.3   & 2.8  &10.4 & 3.2     \\
                             &      &     &       &      &     &         \\
\hline

\end{tabular}

\noindent
$^{a}$: the fluxes are given in units of 10$^{-14}$ erg cm$^{-2}$ s$^{-1}$.\\ 
$^{b}$: fluxes values from CASLEO and CTIO data of moderate spectral
resolution ($\sim$290--300 km s$^{-1}$).\\
$^{c}$: fluxes values from CASLEO data of high spectral resolution
($\sim$50 km s$^{-1}$).\\
$^{d}$: the luminosities are given in units of 10$^{40}$ erg s$^{-1}$.\\ 

\end{table}

 
\begin{table}
\footnotesize \caption{Emission line ratios of NGC\,5514, from long--slit
spectroscopy (medium and high resolution; from CASLEO)}
\label{elr1d}
\begin{tabular}{lcccccl}
\hline \hline

                  &   &         &         &         &         &\\
Regions     & lg[O{\sc iii}]/H$\beta$$^{a}$ & lg[O{\sc i}]/H$\alpha$$^{a}$ 
& lg[N{\sc ii}]/H$\alpha$$^{a}$  & lg[S{\sc ii}s]/H$\alpha$$^{a}$ &
 [S{\sc ii}]/[S{\sc ii}]$^{a}$ & Spectral Type \\
\hline
                  &     &         &         &         &       & \\
{\it Medium Spec. Resolut.} & &   &         &         &       & \\
                  &     &         &         &         &       & \\
Main Nucleus      & --- &  -0.22  &   0.36  &   0.04  &  1.20 & L \\
West Area         & 0.34&  -0.39  &   0.24  &   0.16  &  1.28 & L \\
Sec. Nucleus+H{\sc ii} Reg&0.07&-0.81&-0.35 &  -0.13  &  1.04 & L/H {\sc ii} \\
                  &     &         &         &         &       & \\
{\it High Spec. Resolut.}&    &   &         &         &       & \\
                  &     &         &         &         &       & \\
Main Nucleus      &(0.4)&  -0.25  &   0.35  &  -0.04  &  1.23 & L  \\
West Area         & 0.3 &  -0.36  &   0.20  &   0.13  &  1.25 & L \\
Sec. Nucleus+H{\sc ii} Reg&0.1&-0.63&-0.30  &  -0.15  &  1.12 & L/H {\sc ii} \\
                  &     &         &         &         &       & \\
\hline

\end{tabular}

\noindent
$^{a}$: [O {\sc iii}]${\lambda5007}$; [O {\sc i}]$\lambda6300$; 
[N {\sc ii}]$\lambda6583$; [S {\sc ii}s]$\lambda\lambda$6716+6731;
[O {\sc ii}]${\lambda3727}$;
[S {\sc ii}]/[S {\sc ii}] $\lambda6716$/$\lambda6731$.\\
Column 7: for the spectral type L and L/H {\sc ii} mean: LINERs, and
transition objects, between LINERs and H {\sc ii} regions.\\
The used spectral resolution were 290 and 50 km s$^{-1}$.\\
The values between parenthesis are data with low S/N. 

\end{table}

\clearpage

\begin{table}
\footnotesize \caption{Fluxes of emission lines of NGC\,5514
 (from WHT+INTEGRAL 2D spectroscopy, total values)}
\label{flux2dt}
\begin{tabular}{llccccccccccc}
\hline
\hline

Lines&Comp &  &      &        &Fluxes$^{a}$&&    &      &      &&              &\\
EqW  &     &     &      & Bubble &      &    &      &MNuc  &SNuc  &&H{\sc ii} Reg&\\
FWHM, Lum& &Knot 1&Knot 2a& Knot 2b&Knot 3a&Knot 3b&Knot 4&&&MKnot&EBorder  &WBord\\

\hline

H$\beta\lambda4861$
  & BBC& 3.2 & 1.0 & 0.2 & 0.4 & 0.6 & 1.8 & --- & --- & --- & --- & ---\\
  & RBC& 0.3 & 0.1 & 0.3 & 0.6 & 0.1 & --- & --- & --- & --- & --- & ---\\
  & MC & --- & --- & --- & --- & --- & --- & 0.7 & 1.0 & 1.0 & 0.4 & 0.6\\
  & OF1& --- & --- & --- & --- & --- & --- & --- & --- & --- & 0.1 & 0.1\\

[O{\sc iii}]$\lambda5007$
  & BBC& 4.4 & 1.0 & 0.2 & 0.9 & 0.8 & 2.5 & --- & --- & --- & --- & ---\\
  & RBC& 0.2 & 0.2 & 0.2 & 0.6 & 0.1 & --- & --- & --- & --- & --- & ---\\
  & MC &     & --- & --- & --- & --- & --- & 1.4 & 0.9 & 0.8 & 0.7 & 1.3\\
  & OF1& --- & --- & --- & --- & --- & --- & --- & --- & --- &(0.1)&(0.1)\\

[N {\sc i}]$\lambda5199$
  & BBC& 2.4 & 1.1 &(0.3)& 0.3 & 0.5 & --- & --- & --- & --- & --- & ---\\
  & RBC& --- & --- & 0.6 & --- & --- & --- & --- & --- & --- & --- & ---\\
  & MC & --- & --- & --- & --- & --- & --- & --- & --- & --- & --- & ---\\
  & OF1& --- & --- & --- & --- & --- & --- & --- & --- & --- & --- & ---\\

[O {\sc i}]$\lambda6300$
  & BBC& 7.5 & 2.7 & 0.5 & 0.9 & 0.8 & 3.0 & --- & --- & --- & --- & --- \\
  & RBC& 0.8 & 0.4 & 0.6 & 1.2 & 0.4 & --- & --- & --- & --- & --- & --- \\
  & MC & --- & --- & --- & --- & --- & --- & 2.0 & 1.1 & 0.6 & 0.4 & 1.1 \\
  & OF1& --- & --- & --- & --- & --- & --- & --- & --- & --- & 0.2 & 0.1 \\

H$\alpha\lambda6563$
  & BBC&23.3 & 5.6 & 0.7 & 1.1 & 2.1 & 5.5 & --- & --- & --- & --- & ---\\
  & RBC& 1.3 & 0.7 & 1.0 & 3.7 & 0.8 & --- & --- & --- & --- & --- & ---\\
  & MC & --- & --- & --- & --- & --- & --- & 2.8 & 7.7 & 4.6 & 1.7 & 2.8\\
  & OF1& --- & --- & --- & --- & --- & --- & --- & --- & --- & 1.1 & 0.2\\

[N {\sc ii}]$\lambda6583$
  & BBC&31.6 & 9.4 & 1.0 & 3.1 & 4.0 & 8.1 & --- & --- & --- & --- & ---\\
  & RBC& 1.7 & 1.1 & 1.4 & 3.1 & 1.3 & --- & --- & --- & --- & --- & --- \\
  & MC & --- & --- & --- & --- & --- & --- & 4.8 & 4.1 & 2.4 & 1.8 & 2.7 \\
  & OF1& --- & --- & --- & --- & --- & --- & --- & --- & --- & 1.1 & 0.3 \\

[S {\sc ii}]$\lambda6717$
  & BBC& 9.8 & 2.0 & 0.4 & 0.5 & 1.9 & 3.5 & --- & --- & --- & --- & --- \\
  & RBC& 0.8 &(0.2)& 0.8 & 1.8 &(0.4)& --- & --- & --- & --- & --- & --- \\
  & MC & --- & --- & --- & --- & --- & --- & 1.6 & 2.1 & 0.9 & 0.6 & 1.2 \\
  & OF1& --- & --- & --- & --- & --- & --- & --- & --- & --- &(0.3)& 0.1 \\

[S {\sc ii}]$\lambda6731$
  & BBC& 9.0 & 2.7 & 0.4 &(0.5)& 1.9 & 2.1 & --- & --- & --- & --- & ---\\
  & RBC& 0.8 & 0.2 & 0.7 & 1.7 & 0.4 & --- & --- & --- & --- & --- & ---\\
  & MC & --- & --- & --- & --- & --- & --- & 1.1 & 1.6 & 0.8 & 0.5 & 1.0 \\
  & OF1& --- & --- & --- & --- & --- & --- & --- & --- & --- & 0.2 & 0.1 \\

WR bump
  & MC & 5.7 & 2.1 & 1.4 & 1.2 & 1.4 & --- & --- & --- & --- & --- & ---\\
($\lambda4650$)
  &    &     &     &     &     &     &     &     &     &     &     &  \\

  &    &     &     &     &     &     &     &     &     &     &     &  \\
H$\alpha$/H$\beta$
  & BBC& 7.2 & 9.4 & 3.5 & 2.8 & 3.5 & 3.1 & --- & --- & --- & --- & --- \\
  & RBC& 4.3 & 7.0 & 3.3 & 6.2 & 8.0 & --- & --- & --- & --- & --- & --- \\
  & MC & --- & --- & --- & --- & --- & --- & 4.0 & 7.7 & 4.6 & 4.3 & 4.6 \\
  & OF1& --- & --- & --- & --- & --- & --- & --- & --- & --- &11.0 & 3.0 \\

  &    &     &     &     &     &     &     &     &     &     &     &  \\
EqW H$\alpha$ 
  & BBC& 152 &  47 &  11 &  10 &  25 &  85 & --- & --- & --- & --- & --- \\
(\AA)
  & RBC&   6 &   5 &  15 &  33 &  11 & --- & --- & --- & --- & --- & --- \\
  & MC & --- & --- & --- & --- & --- & --- &   6 &  47 &  79 &  24 &  22\\
  & OF1& --- & --- & --- & --- & --- & --- & --- & --- & --- &  15 &   2\\

FWHM H$\alpha$ 
  & BBC& 350 & 270 & 200 & 250 & 270 & 360 & --- & --- & --- & --- & --- \\
(km/s)
  & RBC& 200 & 260 & 215 & 290 & 230 & --- & --- & --- & --- & --- & --- \\
  & MC & --- & --- & --- & --- & --- & --- & 240 & 150 & 122 & 370 & 270 \\
  & OF1& --- & --- & --- & --- & --- & --- & --- & --- & --- & 330 & 190 \\

  &    &     &     &     &     &     &     &     &     &     &     &  \\
Lum H$\alpha$$^{b}$
  & BBC&26.7 & 6.4 & 0.8 & 1.3 & 2.4 & 6.3 & --- & --- & --- & --- & ---\\
  & RBC& 1.5 & 0.8 & 1.2 & 4.2 & 0.9 & --- & --- & --- & --- & --- & ---\\
  & MC & --- & --- & --- & --- & --- & --- & 3.2 & 8.8 & 5.3 & 2.0 & 3.2\\
  & OF1& --- & --- & --- & --- & --- & --- & --- & --- & --- & 1.3 & 0.3\\

Lum WR 
  & MC & 6.5 & 2.4 & 1.6 & 1.4 & 1.6 & --- & --- & --- & --- & --- & ---\\

\hline

\end{tabular}

\noindent
$^{a}$: the fluxes are given in units of 10$^{-15}$ erg cm$^{-2}$ s$^{-1}$
(from 2D spectroscopy with a resolution of $\sim$100 km s$^{-1}$).\\
$^{b}$: the luminosities are given in units of 10$^{39}$ erg  s$^{-1}$.\\ 
Column 2: emission line components (Comp), where BBC, RBC, MC and OF1 mean
blue bubble component, red bubble component, main component and outflow
system-1, respectively.\\
The values between parenthesis are data with low S/N.\\

\end{table}

\clearpage

\begin{table}
\footnotesize \caption{Emission Line Ratios of NGC\,5514 (2D spectroscopy,
total values)}
\label{elr2dt}
\begin{tabular}{lccccccc}
\hline \hline

               &    &   &       &         &         &        &      \\
Regions        &Comp& lg[O{\sc iii}]/H$\beta$$^{a}$&lg[O{\sc i}]/H$\alpha^{a}$ 
& lg[N{\sc ii}]/H$\alpha$$^{a}$  & lg[S{\sc ii}s]/H$\alpha$$^{a}$ &
[S{\sc ii}]/[S{\sc ii}]$^{a}$ & Spectral Type \\
\hline
               &    &       &       &       &       &       &     \\

{\bf Bubble}   &    &       &       &       &       &       &     \\
{\it Knot 1}   &BBC &  0.14 & -0.49 &  0.13 & -0.09 &  1.09 & L   \\
               &RBC & -0.18 & -0.21 &  0.12 & -0.21 &  1.00 & L   \\
{\it Knot 2a}  &BBC &  0.00 & -0.31 &  0.22 & -0.05 &  0.85 & L   \\
               &RBC &  0.30 & -0.24 &  0.20 & -0.24 & (1.0) & L   \\
{\it Knot 2b}  &BBC &  0.00 & -0.20 &  0.16 &  0.06 & (1.0) & L   \\
               &RBC & -0.17 & -0.22 &  0.15 &  0.18 &  1.14 & L   \\
{\it Knot 3a}  &BBC &  0.35 & -0.09 &  0.45 &  0.00 &  1.00 & L   \\
               &RBC &  0.00 & -0.49 & -0.08 & -0.02 &  1.06 & L   \\
{\it Knot 3b}  &BBC &  0.13 & -0.34 &  0.30 &  0.24 &  1.11 & L   \\
               &RBC &  0.00 & -0.30 &  0.24 &  0.00 &  1.00 & L   \\
{\it Knot 4}   &BBC &  0.14 & -0.26 &  0.17 &  0.01 &  1.66 & L   \\
               &    &       &       &       &       &       &     \\
{\bf Main Nucleus}
               &MC  &  0.30 & -0.15 &  0.23 & -0.02 &  1.45 &  L  \\

               &    &       &       &       &       &       &     \\

{\bf Second Nucleus}
               &MC  & -0.05 & -0.85 & -0.27 & -0.32 &  1.31 &L/H {\sc ii}  \\
               &    &       &       &       &       &       &     \\
{\bf H {\sc ii} Region Complex}
               &    &       &       &       &       &       &     \\
{\it Main Knot (H {\sc ii} Reg)}
               &MC  & -0.10 & -0.88 & -0.28 & -0.43 &  1.13 &L/H {\sc ii}  \\
{\it East-Border (H {\sc ii} Reg)}
               &MC  &  0.24 & -0.63 &  0.03 & -0.19 &  1.20 & L   \\
               &OF1 &  0.00 & -0.74 &  0.00 & -0.34 &  1.50 & L   \\
{\it West-Border (H {\sc ii} Reg)}
               &MC  &  0.33 & -0.40 & -0.02 & -0.11 &  1.20 & L   \\
               &OF1 &  0.00 & -0.30 &  0.18 &  0.00 &  1.00 & L   \\
               &    &       &       &       &       &       &     \\

\hline

\end{tabular}

\noindent
$^{a}$: [O {\sc iii}]$\lambda5007$; [O {\sc i}]$\lambda6300$; 
[N {\sc ii}]$\lambda6583$; [S {\sc ii}s]$\lambda\lambda$6716+6731;
[S {\sc ii}]/[S {\sc ii}] $\lambda6716$/$\lambda6731$.\\
Column 2: emission line components (Comp), where BBC, RBC, MC and OF1 mean
blue bubble component, red bubble component, main component and outflow
system-1, respectively.\\
Column 8: for the spectral type, L and L/H {\sc ii} mean: LINERs, and
transition objects, between LINERs and H {\sc ii} regions.\\
The values between parentheses are data with low S/N.

\end{table}

\clearpage

\begin{table}
\footnotesize \caption{Fluxes of Emission Lines of NGC\,5514
 (from WHT+INTEGRAL 2D spectroscopy, central fibre values)}
\label{flux2df}
\begin{tabular}{llccccccccccl}
\hline
\hline

Lines&Comp &     &      &   &Fluxes$^{a}$&&     &      &      &&              & \\
EqW  &     &     &      & Bubble &      &&      &MNuc &SNuc &&H{\sc ii} Reg& \\
FWHM &     &Knot 1&Knot 2a& Knot 2b&Knot 3a&Knot 3b&Knot 4&    &  &MKnot&EBorder&WBord\\
 &&P1-f126&P1-f101&P1-f47&P1-f96&P1-f103&P1-f123&P2-f138&P3-f93&P3-f68&P3-f48&P3-f101\\

\hline

H$\beta\lambda4861$
  & BBC& 0.5 & 0.3 & 0.1 & 0.2 & 0.3 & 0.5 & --- & --- & --- & --- & --- \\
  & RBC& 0.1 &     & 0.1 & 0.2 & --- & --- & --- & --- & --- & --- & --- \\
  & MC & --- & --- & --- & --- & --- & --- & 0.3 & 0.3 & 0.4 & 0.2 & 0.2 \\
  & OF1& --- & --- & --- & --- & --- & --- & --- & --- & --- & 0.1 & --- \\

[O{\sc iii}]$\lambda5007$
  & BBC& 0.6 & 0.4 & 0.1 & 0.3 & 0.5 & 0.7 & --- & --- & --- & --- & --- \\
  & RBC& 0.1 &     & 0.2 & 0.3 & --- & --- & --- & --- & --- & --- & --- \\
  & MC & --- & --- & --- & --- & --- & --- & 0.5 & 0.4 & 0.4 & 0.3 & 0.3 \\
  & OF1& --- & --- & --- & --- & --- & --- & --- & --- & --- &(0.1)& --- \\

[N {\sc i}]$\lambda5199$
  & BBC& 0.4 & 0.3 &(0.2)& 0.2 &(0.2)& --- & --- & --- & --- & --- & --- \\
  & RBC&(0.2)& --- &(0.2)& --- & --- & --- & --- & --- & --- & --- & --- \\
  & MC & --- & --- & --- & --- & --- & --- & --- & --- & --- & --- & --- \\
  & OF1& --- & --- & --- & --- & --- & --- & --- & --- & --- & --- & --- \\

[O {\sc i}]$\lambda6300$
  & BBC& 0.6 & 0.7 & 0.1 & 0.3 & 0.4 & 1.0 & --- & --- & --- & --- & --- \\
  & RBC& 0.3 &(0.1)& 0.4 & 0.5 & 0.2 & --- & --- & --- & --- & --- & --- \\
  & MC & --- & --- & --- & --- & --- & --- & 0.9 & 0.4 & 0.4 & 0.4 & 0.5 \\
  & OF1& --- & --- & --- & --- & --- & --- & --- & --- & --- & 0.1 & --- \\

H$\alpha\lambda6563$
  & BBC& 2.3 & 1.7 & 0.3 & 1.4 & 1.0 & 1.7 & --- & --- & --- & --- & --- \\
  & RBC& 1.0 & 0.2 & 0.7 & 0.8 & 0.5 & --- & --- & --- & --- & --- & --- \\
  & MC & --- & --- & --- & --- & --- & --- & 1.2 & 1.9 & 3.1 & 0.8 & 1.0 \\
  & OF1& --- & --- & --- & --- & --- & --- & --- & --- & --- & 0.3 & --- \\

[N {\sc ii}]$\lambda6583$
  & BBC& 2.5 & 2.7 & 0.4 & 1.6 & 2.0 & 2.1 & --- & --- & --- & --- & --- \\
  & RBC& 1.4 & 0.3 & 0.9 & 0.7 & 0.6 & --- & --- & --- & --- & --- & --- \\
  & MC & --- & --- & --- & --- & --- & --- & 1.8 & 1.4 & 1.6 & 0.7 & 0.8 \\
  & OF1& --- & --- & --- & --- & --- & --- & --- & --- & --- & 0.3 & --- \\

[S {\sc ii}]$\lambda6717$
  & BBC& 0.7 & 0.7 &(0.1)& 0.8 & 0.9 & 1.1 & --- & --- & --- & --- & --- \\
  & RBC& 0.5 &(0.1)& 0.5 & 0.4 & 0.2 & --- & --- & --- & --- & --- & --- \\
  & MC & --- & --- & --- & --- & --- & --- & 0.6 & 0.7 & 0.5 & 0.4 & 0.5 \\
  & OF1& --- & --- & --- & --- & --- & --- & --- & --- & --- & 0.2 & --- \\

[S {\sc ii}]$\lambda6731$
  & BBC& 0.8 & 0.8 & 0.1 &(0.8)& 0.9 & 1.6 & --- & --- & --- & --- & --- \\
  & RBC& 0.4 & 0.2 & 0.4 & 0.3 & 0.2 & --- & --- & --- & --- & --- & --- \\
  & MC & --- & --- & --- & --- & --- & --- & 0.5 & 0.5 & 0.5 & 0.4 & 0.4 \\
  & OF1& --- & --- & --- & --- & --- & --- & --- & --- & --- & 0.2 & --- \\

  &    &     &     &     &     &     &     &     &     &     &     &     \\
H$\alpha$/H$\beta$
  & BBC& 4.6 & 5.7 & 3.0 & 7.0 & 3.3 & 3.4 & --- & --- & --- & --- & --- \\
  & RBC&10.0 & --- & 7.0 & 4.0 & --- & 1.8 & --- & --- & --- & --- & --- \\
  & MC & --- & --- & --- & --- & --- & --- & 4.0 & 6.3 & 7.7 & 4.0 & 5.0 \\
  & OF1& --- & --- & --- & --- & --- & --- & --- & --- & --- & 3.0 & --- \\

  &    &     &     &     &     &     &     &     &     &     &     &     \\
EqW H$\alpha$ 
  & BBC& 400 &  66 &   8 &  47 &  29 &  81 & --- & --- & --- & --- & --- \\
(\AA)
  & RBC& 110 &   8 &  27 &  24 &  12 & --- & --- & --- & --- & --- & --- \\
  & MC & --- & --- & --- & --- & --- & --- &   7 &  32 & 104 &  21 &  22 \\
  & OF1& --- & --- & --- & --- & --- & --- & --- & --- & --- &   5 & --- \\

  &    &     &     &     &     &     &     &     &     &     &     &     \\
FWHM H$\alpha$ 
  & BBC& 325 & 240 & 200 & 380 & 210 & 400 & --- & --- & --- & --- & --- \\
(km/s)
  & RBC& 275 & 240 & 280 & 252 & 200 & --- & --- & --- & --- & --- & --- \\
  & MC & --- & --- & --- & --- & --- & --- & 235 & 190 & 110 & 320 & 250 \\
  & OF1& --- & --- & --- & --- & --- & --- & --- & --- & --- & 180 & --- \\

\hline

\end{tabular}

\noindent
$^{a}$: the fluxes are given in units of 10$^{-15}$ erg cm$^{-2}$ s$^{-1}$
(from 2D spectroscopy with a resolution of $\sim$100 km s$^{-1}$).\\
Column 2: emission line components (Comp), where BBC, RBC, MC and OF1 mean
blue bubble component, red bubble component, main component and outflow
system-1, respectively.\\
P1, P2 and P3 mean Position 1, 2, and 3, of the fibre bundle.\\
The values between parentheses are data with low S/N.\\

\end{table}

\clearpage

\begin{table}
\footnotesize \caption{Emission Line Ratios of NGC\,5514 (2D spectroscopy,
 individual/central fibres values)}
\label{elr2df}
\begin{tabular}{lccccccc}
\hline \hline

               &    &   &       &         &         &        &      \\
Regions        &Comp& lg[O{\sc iii}]/H$\beta$$^{a}$&lg[O{\sc i}]/H$\alpha^{a}$ 
& lg[N{\sc ii}]/H$\alpha$$^{a}$  & lg[S{\sc ii}s]/H$\alpha$$^{a}$ &
 [S{\sc ii}]/[S{\sc ii}]$^{a}$ & Spectral Type \\
\hline
               &    &       &       &       &       &       &       \\

{\bf Bubble}   &    &       &       &       &       &       &       \\
{\it Knot 1}   &    &       &       &       &       &       &       \\
Pos.1-fibre 126&BBC &  0.08 & -0.58 &  0.04 & -0.19 &  0.88 & L     \\
               &RBC &  0.00 & -0.52 &  0.15 & -0.05 &  1.25 & L     \\
{\it Knot 2a}  &    &       &       &       &       &       &       \\
Pos.1-fibre 101&BBC &  0.12 & -0.39 &  0.20 & -0.05 &  0.88 & L     \\
               &RBC &  ---  & -0.30 &  0.18 &  0.18 & (0.5) & L     \\
{\it Knot 2b}  &    &       &       &       &       &       &       \\
Pos.1-fibre 47 &BBC &  0.00 & -0.30 &  0.17 &  0.00 &  1.00 & L     \\
               &RBC &  0.30 & -0.20 &  0.11 &  0.11 &  1.25 & L     \\
{\it Knot 3a}  &    &       &       &       &       &       &       \\
Pos.1-fibre 096&BBC &  0.30 & -0.67 &  0.06 &  0.06 &  1.00 & L     \\
               &RBC &  0.18 & -0.20 & -0.06 & -0.12 &  1.00 & L     \\
{\it Knot 3b}  &    &       &       &       &       &       &       \\
Pos.1-fibre 103&BBC &  0.22 & -0.40 &  0.30 &  0.25 &  1.00 & L     \\
               &RBC &  ---  & -0.40 &  0.08 & -0.10 &  1.00 & L     \\
{\it Knot 4}   &    &       &       &       &       &       &       \\
Pos.1-fibre 123&BBC &  0.15 & -0.23 &  0.09 &  0.00 &  1.21 & L     \\
               &    &       &       &       &       &       &       \\
{\bf Main Nucleus}& &       &       &       &       &       &       \\
Pos. 2-fibre 138&MC &  0.10 & -0.13 &  0.18 & -0.04 &  1.20 & L     \\

               &    &       &       &       &       &       &       \\

{\bf Second Nucleus}& &     &       &       &       &       &       \\
Pos. 3-fibre 093&MC &  0.13 & -0.70 & -0.13 & -0.20 &  1.40 &L      \\
               &    &       &       &       &       &       &       \\
{\bf H {\sc ii} Region Complex}&&& &       &       &       &       \\
{\it Main Knot (H {\sc ii} Reg)}& & & &     &       &       &       \\
Pos. 3-fibre68 &MC  &  0.00 & -0.89 & -0.29 & -0.49 &  1.07 &L/H {\sc ii} \\
{\it East-Border (H {\sc ii} Reg)}&&&&      &       &       &       \\
Pos.3-fibre 048&MC  &  0.18 & -0.30 & -0.06 &  0.00 &  1.00 & L     \\
               &OF1 &  0.00 & -0.30 &  0.14 &  0.15 &  1.00 & L     \\
{\it West-Border (H {\sc ii} Reg)}&&&&      &       &       &       \\
Pos.3-fibre 101&MC  &  0.30 & -0.47 &  0.00 &  0.12 &  1.25 & L     \\

               &    &       &       &       &       &       &       \\

\hline

\end{tabular}

\noindent
$^{a}$: [O {\sc iii}]$\lambda5007$; [O {\sc i}]$\lambda6300$; 
[N {\sc ii}]$\lambda6583$; [S {\sc ii}s]$\lambda\lambda$6716+6731;
[S {\sc ii}]/[S {\sc ii}] $\lambda6716$/$\lambda6731$.\\
Column 2: emission line components (Comp), where BBC, RBC, MC and OF1 mean
blue bubble component, red bubble component, main component and outflow
system-1, respectively.\\
Column 8: for the spectral type,  L and L/H {\sc ii} mean:  LINERs, and
transition objects, between LINERs and H {\sc ii} regions.\\
Pos.1, 2 and 3, mean Position 1, 2, and 3, of the fibre bundle.\\
The values between parentheses are data with low S/N.

\end{table}

\clearpage

\begin{table}
\footnotesize \caption{IR Mergers and IR QSOs/AGN with low and
high/extreme velocity outflow} \label{gwmergersqso}
\begin{tabular}{lcccccccccl}
\hline \hline

Object       & $V_{\rm OF\,1}$ & $V_{\rm OF\,2}$ & $z$&$L_{\rm FIR}$&$L_{\rm
IR}$/$L_{B}$&Nuclear &Morph&RF&OF &OF  \\
&km\,s$^{-1}$ &km\,s$^{-1}$ &    &$\log L/L_{\odot}$ &&activity&Type
&&Type &Reference\\
\hline

{\it Low velocity OF}&          &        &          &     &         &     &    & & &\\

Arp 220       & $-$450    &  ---   & 0.01825  &12.18& 87      &L+SB & PM &---&EL&Heckman et al. (1987)\\
IRAS00182-7112& $-$450    &  ---   & 0.3270   &12.90&(200)    &L+SB &(M) &---&EL&Heckman et al. (1990)\\
IRAS03250+1606& $-$431    &  ---   &  0.129   &12.06& ---     &L    & M  &---&AL&Rupke et al. (2002)\\
IRAS03514+1546& $-$200    &  ---   & 0.02100  &11.10& ---     &  SB & M  &---&AL&Heckman et al. (2000)\\
IRAS09039+0503& $-$656    &  ---   &  0.125   &12.07& ---     &L    & M  &---&AL&Rupke et al. (2002)\\
IRAS11387+4116& $-$511    &  ---   &  0.149   &12.18& ---     &  SB & OM &---&AL&"\\
IRAS23128-5919& $-$300    &  ---   & 0.04490  &12.60&  8      &L+SB & M  &---&EL&L\'{\i}pari et al. (2003) \\
Mrk 266       & $-$300    &  ---   & 0.02900  &11.37&  8      &  SB & PM &---&EL&Wang et al. (1997)\\
Mrk 273       & $-$600    &  ---   & 0.03850  &12.14& 36      &L+SB & PM &---&EL&Colina et al. (1999)\\
NGC 1614      & $-$400    &  ---   & 0.01550  &11.61& 18      &L+SB & M  &---&EL&Ulrich (1972)\\
NGC 2623      & $-$405    &  ---   & 0.01846  &11.55& 17      &L+SB & M  &---&EL&This paper\\
NGC 3256      & $-$370    &  ---   & 0.00940  &11.57&  9      &  SB & MM &---&EL&L\'{\i}pari et al. (2000)\\
NGC 3690      & $-$300    &  ---   & 0.01025  &11.91& 23      &  SB & PM &---&EL&Heckman et al. (1990)\\
NGC\,4039     & $-$365    &  ---   & 0.00560  &10.99&  6      &L+SB & PM &---&EL&L\'{\i}pari et al. (2003)\\
NGC\,5514     & $-$320    &  ---   & 0.02453  &10.70&  1      &L+SB & PM &---&EL&This programme\\
{\it High/Extreme OF}&          &        &          &     &         &     &    &   &  &\\

IRAS01003-2238& $-$770    &$-$1520 &  0.1180  &12.24& 67      &QSO+SB&OM &---&EL&L\'{\i}pari et al. (2003)\\
IRAS05024-1941& $-$1676   &  ---   &  0.192   &12.43& ---     &S2    & M &---&AL&Rupke et al. (2002)\\
IRAS05024-1941& $-$1150   &$-$2127 &  0.192   &12.43& ---     &S2    & M &---&EL&This paper\\
IRAS05189-2524& $-$849    &  ---   &  0.042   &12.07& ---     &S2    & M &---&AL&Rupke et al. (2002)\\
IRAS10378+1108& $-$1517   &  ---   &  0.136   &12.26& ---     &L     & M &---&AL&"\\
IRAS11119+3257& $-$1300  &($-$2120)&  0.1873  &12.58& ---     &S1+SB & M &---&EL&L\'{\i}pari et al. (2003)\\
IRAS13218+0552& $-$1800  &($-$3438)&  0.2048  &12.63& 96      &QSO+SB&OM &---&EL&"\\
IRAS14394+5332& $-$880   &($-$1650)&  0.1050  &12.04& ---     &S2    &MM &---&EL&" \\
IRAS15130-1958& $-$780   &($-$1200)&  0.1093  &12.09& ---     &S2    & M &---&EL&"\\
IRAS15462-0450& $-$1000  &($-$1760)&  0.1001  &12.16& ---     &S1    & M &---&EL&"\\
IRAS19254-7245& $-$800    &  ---   &  0.05970 &12.04& 30      &QSO+SB& PM&---&EL&Colina et al. (1991)\\
Mrk\,231      & $-$1000   &  ---   &  0.04220 &12.53& 32      &QSO+SB& M &1.83&EL&L\'{\i}pari et al. (1994)\\
Mrk\,231      & $-$1100   &  ---   &  0.04220 &12.53& 32      &QSO+SB& M &1.83&OSM&Zheng et al (2002)\\
NGC\,3079     & $-$1600   &  ---   &  0.00400 &10.49&  2      & L+SB &Sp &---&EL&Heckman et al. (1990)\\
NGC\,6240     & $-$930    &  ---   &  0.02425 &11.83& 15      & L+SB &PM &---&EL&"\\
{\it IR QSOoffset-OF}&          &        &          &     &         &      &   &  & &\\
IRAS00275-2859&  $-$730   &  ---   & 0.2792   &12.64&  9      &QSO+SB&PM &1.47&OSM&Zheng et al (2002)\\
IRAS00509+1225& $-$1110   &  ---   & 0.060    &11.87& 20      &QSO+SB&PM &1.40&OSM&This paper\\
IRASZ01373+0604&$-$1660   &(-3100) & 0.3964   &12.30&  3      &QSO+SB&-- &2.33&OSM&"\\
IRAS02054+0835& $-$1355   &  ---   & 0.345    &12.97& ---     &QSO   &-- &2.42&OSM&Zheng et al (2002)\\
IRAS02065+4705&  $-$500   &  ---   & 0.132    &12.27& ---     &QSO   &-- &1.65&OSM&"\\
IRAS04415+1215&  $-$875   &  ---   & 0.089    &12.41&         &QSO   &-- &1.74&OSM&"\\
IRAS04505-2958& $-$1700   &  ---   & 0.2863   &12.55& 20      &QSO+SB&PM &1.33&OSM&This paper\\
IRAS06269-0543&  $-$550   &  ---   & 0.117    &12.49& ---     &QSO   &-- &0.57&OSM&Zheng et al (2002)\\
IRAS07598+6508& $-$2030   &  ---   & 0.1483   &12.45&  5      &QSO+SB& M &2.75&OSM&"\\
IRAS07598+6508& $-$1920   &  ---   & 0.1483   &12.45&  5      &QSO+SB& M &2.75&OSM&This paper\\
IRAS09427+1929& $-$1640   &  ---   & 0.284    &12.61& ---     &QSO   &-- &2.98&OSM&Zheng et al (2002)\\
IRAS10026+4347&  $-$650   &  ---   & 0.178    &12.20& ---     &QSO   &-- &2.07&OSM&"\\
IRAS11119+3527&  $-$950   &  ---   & 0.189    &12.64& ---     &QSO   & M &1.12&OSM&"\\
IRASZ11598-0112& $-$970   &  ---   & 0.151    &11.91& ---     &QSO   &-- &1.71&OSM&"\\
IRAS13305-1739&  $-$730   &  ---   & 0.148    &12.21& ---     &S2    & OM&--- &OSM&This paper\\
IRAS13342+3932&  $-$920   &  ---   & 0.179    &12.49& ---     &QSO   &-- &0.73&OSM&Zheng et al (2002)\\
IRAS13451+1232& $-$1210   &  ---   & 0.122    &12.28& ---     &S2    & PM&--- &OSM&This paper\\
IRAS14026+4341& $-$1500   &  ---   & 0.320    &12.55& 40      &QSO   & M &1.12&OSM&"\\
IRAS15462-0450& $-$1110   &  ---   & 0.101    &12.35& ---     &QSO   & M &1.32&OSM&Zheng et al (2002)\\
IRAS17002+5153& $-$1050   &  ---   & 0.2923   &12.58&  5      &QSO+SB&PM &1.56&OSM&This paper\\
IRAS18508-7815& $-$1250   &  ---   & 0.162    &12.0 &  8      &QSO   & M &2.40&OSM&"\\
IRAS20036-1547&  $-$400   &  ---   & 0.193    &12.70& ---     &QSO   &-- &2.74&OSM&Zheng et al (2002)\\
IRAS20520-2329&  $-$660   &  ---   & 0.206    &12.52& ---     &QSO   &-- &2.02&OSM&"\\
IRAS21219-1757&  $-$460   &  ---   & 0.110    &11.98& 89      &QSO   &OM &1.82&OSM&"\\
IRAS22419-6049&  $-$390   &  ---   & 0.1133   &11.30& ---     &QSO+SB& M &--- &OSM&This paper\\
IRAS23389+0300& $-$1640   &  ---   & 0.145    &12.09& ---     &S2    & PM&--- &OSM&"\\

\hline

\end{tabular}

\end{table}

\clearpage

Notes.
Cols.\ 2 and 3: OF values obtained from the references included in Col.\ 9.
Values between parentheses are possible detections.
[O {\sc iii}] MC indicates emission line                                          
[O {\sc iii}]\,$\lambda$5007 with multiple components.

Col.\ 5: The L$_{\rm IR}$ were obtained for [8--1000 $\mu$m], using the
relation given by Sanders \& Mirabel (1996).

Col.\ 7: The properties of the nuclear activity were obtained from Veilleux
et al. (1999), Canalizo \& Stockton (2001) and L\'{\i}pari et al.\ (2003). S1:
Seyfert 1, S2: Seyfert 2, L: Liners and SB: starburst.

Col.\ 8: For the morphological or interaction type we used the classification
criteria of Veilleux, Kim \& Sanders (2002b) and Sourace (1998); PM:
pre--merger, M: merger, OM: old merger, MM: multiple merger, Sp: spirals.

Col.\ 9: RF is the ratio of the emission line
Fe {\sc ii}$\lambda$4570/H$\beta$

Col.\ 10: EL, AL and OSM indicate OF derived from emission, absorption lines,
and offset emission line method (see the text), respectively.


\clearpage

\begin{figure*}
\vspace{12.0 cm}
\begin{tabular}{c}
\includegraphics{fig1az.ps} \cr
\includegraphics{fig1bz.ps} \cr
\end{tabular}
\vspace{6.0 cm}
\caption {
CASLEO $V$  image and contour of the whole IR
pre--merger NGC\,5514. North is to the top, east is to the left.
}
\label{wholemer}
\end{figure*}

\clearpage
\begin{figure*}
\vspace{12.0 cm}
\begin{tabular}{cc}
\includegraphics{fig2az.ps}&
\includegraphics{fig2bz.ps} \cr
\includegraphics{fig2cz.ps}&
\includegraphics{fig2dz.ps} \cr
\end{tabular}
\vspace{6.0 cm}
\caption {
CASLEO optical broad--band contour map images of NGC\,5514  (through the
filters U, B, V and I). North is up, and the east to the left.
}
\label{caimages}
\end{figure*}

\clearpage
\begin{figure*}
\vspace{12.0 cm}
\begin{tabular}{c}
\includegraphics{fig3az.ps} \cr
\includegraphics{fig3bz.ps} \cr
\includegraphics{fig3cz.ps}  \cr
\end{tabular}
\vspace{8.0 cm}
\caption {
2MASS near IR contour map images of NGC\,5514  (through the
filters J, H and K$_S$). North is to the top, east is to the left.
}
\label{2mimages}
\end{figure*}

\clearpage
\begin{figure*}
\vspace{12.0 cm}
\begin{tabular}{c}
\includegraphics{fig4z.ps} \cr
\end{tabular}
\vspace{6.0 cm}
\caption {
Palomar 103aO ($\sim$ B)  deep plate image of NGC\,5514. The arrows show
the position of 3 faint external filaments (see the text).
North is up, and the east to the left.
}
\label{palimage}
\end{figure*}

\clearpage
\begin{figure*}
\vspace{12.0 cm}
\begin{tabular}{c}
\includegraphics{fig5az.ps} \cr
\includegraphics{fig5bz.ps} \cr
\end{tabular}
\vspace{6.0 cm}
\caption {
CASLEO colour $(B - I)$ broad--band image and contour map of NGC\,5514.
North is up, and the east to the left.
}
\label{redening}
\end{figure*}

\clearpage
\begin{figure*}
\vspace{12.0 cm}
\begin{tabular}{c}
\includegraphics{fig6az.ps} \cr
\includegraphics{fig6bz.ps} \cr
\end{tabular}
\vspace{6.0 cm}
\caption {
{\itshape 2MASS\/ K$_S$} surface brightness (mag arcsec$^{-2}$) plotted
against the fourth root of the radius, for the main (a) and the second (b)
galaxies, of the original systems that collide.
}
\label{perfill}
\end{figure*}


\clearpage
\begin{figure*}
\vspace{12.0 cm}
\begin{tabular}{c}
\includegraphics{fig7az.ps} \cr
\includegraphics{fig7bz.ps} \cr
\includegraphics{fig7cz.ps} \cr
\end{tabular}
\vspace{8.0 cm}
\caption {
CASLEO long--slit spectra (of medium spectral resolution: 290 k s$^{-1}$)
for the 3 main emission regions,  in NGC\,5514.
The scales of flux are given in units of [erg $\times$ cm$^{-2}$ $\times$
s$^{-1}$ $\times$ \AA$^{-1}$] and wavelength in [$\AA$].
}
\label{1dspec1}
\end{figure*}

\clearpage
\begin{figure*}
\vspace{12.0 cm}
\begin{tabular}{c}
\includegraphics{fig8az.ps} \cr
\includegraphics{fig8bz.ps} \cr
\includegraphics{fig8cz.ps} \cr
\end{tabular}
\vspace{8.0 cm}
\caption {
CASLEO long--slit spectra (of high spectral resolution: 50 k s$^{-1}$) for
the 3 main emission regions  of NGC\,5514.
The scales of flux are given in units of [erg $\times$ cm$^{-2}$ $\times$
s$^{-1}$ $\times$ \AA$^{-1}$] and wavelength in [$\AA$].
}
\label{1dspec2}
\end{figure*}


\clearpage
\begin{figure*}   
\vspace{12.0 cm}
\begin{tabular}{c}
\includegraphics{fig9az.ps} \cr
\includegraphics{fig9bz.ps} \cr
\end{tabular}
\vspace{6.0 cm}
\caption {
WHT+INTEGRAL mosaics maps of the continuum$\lambda\lambda$6540--6640 (a)
and H$\alpha$+[N {\sc ii}] emission (b).
The position of the main nucleus was defined as the zero point (0,0).
North is up, and the east to the left.
}
\label{whtmosaic12}
\end{figure*}

\clearpage
\begin{figure*}
\vspace{12.0 cm}
\begin{tabular}{c}
\includegraphics{fig10z.ps}\cr
\end{tabular}
\vspace{6.0 cm} \caption {
Superposition of the bubble H$\alpha$+[N {\sc ii}] map and 2D Integral
spectra of NGC\,5514
(the spectra are shown at the wavelength region of H$\alpha$ + [N\,{\sc ii}]).
The number on each spectra indicate the fibre number.
North is up, and the east to the left.
}
\label{idab}
\end{figure*}

\clearpage
\begin{figure*}
\vspace{12.0 cm}
\begin{tabular}{cc}
\includegraphics{fig11az.ps}& 
\includegraphics{fig11bz.ps} \cr
\includegraphics{fig11cz.ps}&
\includegraphics{fig11dz.ps} \cr
\includegraphics{fig11ez.ps}&
\includegraphics{fig11fz.ps} \cr
\end{tabular}
\vspace{8.0 cm}
\caption {
WHT+Integral maps of the supergiant bubble in NGC\,5514 (position 1) for
(a) narrow continuum
adjacent to H$\alpha$;  (b), (c), (d), (e) and (f) pure H$\alpha$,
[N\,{\sc ii}], [S\,{\sc ii}], [O\,{\sc i}], [O\,{\sc iii}] emission line,
respectively (the continuum was subtracted).
The crosses show the position of the knots.
The position of the centre of the fibre bundle was defined as the zero (0,0)
value.
North is up, and the east to the left.
}
\label{bubblei}
\end{figure*}

\clearpage
\begin{figure*}
\vspace{12.0 cm}
\begin{tabular}{cc}
\includegraphics{fig12az.ps}& 
\includegraphics{fig12bz.ps} \cr
\includegraphics{fig12cz.ps}&
\includegraphics{fig12dz.ps} \cr
\includegraphics{fig12ez.ps}&
\includegraphics{fig12fz.ps} \cr
\end{tabular}
\vspace{8.0 cm}
\caption {
WHT+Integral 2D spectra of the main knots of the bubble.
In the blue and red wavelength regions.
}
\label{bubbles}
\end{figure*}

\clearpage
\begin{figure*}
\vspace{12.0 cm}
\begin{tabular}{cc}
\includegraphics{fig12gz.ps}& 
\includegraphics{fig12hz.ps} \cr
\includegraphics{fig12iz.ps}&
\includegraphics{fig12jz.ps} \cr
\includegraphics{fig12kz.ps}&
\includegraphics{fig12lz.ps} \cr
\end{tabular}
\vspace{8.0 cm}
\addtocounter{figure}{-1}
\caption {
Continuation.
}
\label{bubblesc}
\end{figure*}

\clearpage
\begin{figure*}
\vspace{12.0 cm}
\begin{tabular}{cc}
\includegraphics{fig13az.ps}&
\includegraphics{fig13bz.ps} \cr
\includegraphics{fig13cz.ps}&
\includegraphics{fig13dz.ps} \cr
\end{tabular}
\vspace{6.0 cm}
\caption {
WHT+Integral maps of the nuclei region of NGC\,5514 (position 2) for
(a) narrow continuum
adjacent to H$\alpha$;  (b), (c) and (d) pure [N\,{\sc ii}], H$\alpha$, 
and [S\,{\sc ii}] emission line, respectively
(the continuum was subtracted).
The position of the centre of the fibre bundle was defined as the zero (0,0)
value.
North is up, and the east to the left.
}
\label{nucleii}
\end{figure*}

\clearpage
\begin{figure*}
\vspace{12.0 cm}
\begin{tabular}{cc}
\includegraphics{fig14az.ps}&
\includegraphics{fig14bz.ps} \cr
\includegraphics{fig14cz.ps}&
\includegraphics{fig14dz.ps} \cr
\end{tabular}
\vspace{6.0 cm}
\caption {
WHT+Integral spectra of the nuclei region in NGC\,5514 (position 2).
In the blue and red wavelength regions.
}
\label{nucleis}
\end{figure*}

\clearpage
\begin{figure*}
\vspace{12.0 cm}
\begin{tabular}{cc}
\includegraphics{fig15az.ps}& 
\includegraphics{fig15bz.ps} \cr
\includegraphics{fig15cz.ps}&
\includegraphics{fig15dz.ps} \cr
\end{tabular}
\vspace{6.0 cm}
\caption {
WHT+Integral maps of the  complex H {\sc ii} region in NGC\,5514 (position 3)
for (a) narrow continuum
adjacent to H$\alpha$;  (b), (c) and (d) pure H$\alpha$, [N\,{\sc ii}]
and [S\,{\sc ii}] emission line, respectively
(the continuum was subtracted).
The position of the centre of the fibre bundle was defined as the zero (0,0)
value.
North is up, and the east to the left.
}
\label{h2i}
\end{figure*}

\clearpage
\begin{figure*}
\vspace{12.0 cm}
\begin{tabular}{cc}
\includegraphics{fig16az.ps}&
\includegraphics{fig16bz.ps} \cr
\includegraphics{fig16cz.ps}&
\includegraphics{fig16dz.ps} \cr
\end{tabular}
\vspace{6.0 cm}
\caption {
WHT+Integral spectra of the complex of H {\sc ii} region in NGC\,5514
(position 3).
}
\label{h2s}
\end{figure*}


\clearpage
\begin{figure*}
\vspace{12.0 cm}
\begin{tabular}{cc}
\includegraphics{fig17az.ps}&
\includegraphics{fig17bz.ps} \cr
\includegraphics{fig17cz.ps}&
\includegraphics{fig17dz.ps} \cr
\includegraphics{fig17ez.ps}&
\includegraphics{fig17fz.ps} \cr
\end{tabular}
\vspace{6.0 cm}
\caption {
WHT + Integral spectra for individual fibres
(in the wavelength region of the H$\alpha$ + [N\,{\sc ii}] emission line),
showing examples of multiple emission line components.
The scales of flux are given in units of [erg $\times$ cm$^{-2}$ $\times$
s$^{-1}$ $\times$ \AA$^{-1}$].
} \label{mcomp1}
\end{figure*}

\clearpage
\begin{figure*}
\vspace{12.0 cm}
\begin{tabular}{c}
\includegraphics{fig18z.ps}\cr
\end{tabular}
\vspace{6.0 cm} 
\caption {
WHT + Integral map of the Red Bubble Components (RBC), for
[N {\sc ii}]$\lambda$6583 emission line (at the position 1).
The centre of the fibre bundle define the zero (0,0) position.
North is up, and the east to the left.
}
\label{mcomp2}
\end{figure*}


\clearpage
\begin{figure*}
\vspace{12.0 cm}
\begin{tabular}{c}
\includegraphics{fig19z.ps} \cr
\end{tabular}
\vspace{6.0 cm}
\caption {
WHT+Integral velocity field mosaics, for the [N {\sc ii}]$\lambda$6583
emission line, of the ionized gas.
The position of the main nucleus was defined as the zero (0,0) point.
North is up, and the east to the left.
The scales and ranges of velocity are given in units of [km $\times$ s$^{-1}$].
}
\label{whtmosaic34}
\end{figure*}

\clearpage
\begin{figure*}
\vspace{12.0 cm}
\begin{tabular}{cc}
\includegraphics{fig20az.ps}& 
\includegraphics{fig20bz.ps} \cr
\includegraphics{fig20cz.ps}&
\includegraphics{fig20dz.ps} \cr
\end{tabular}
\vspace{6.0 cm}
\caption {
Contour of WHT + Integral  emission line velocity fields of the ionized gas,
at position 1 (i.e. centred in the bubble), for:
(a) [N\,{\sc ii}]$\lambda$6583;
(b) [S\,{\sc ii}]$\lambda$6731;
(c) H$\alpha$;
(d) [O\,{\sc iii}]$\lambda$5007.
The dashed lines show negative values.
The centre of the fibre bundle define the zero (0,0) position.
The ranges of the contours are:
(a) from -630 to 560, step 70 km s$^{-1}$;
(b) from -515 to 600, step 70 km s$^{-1}$;
(c) from -630 to 650, step 80 km s$^{-1}$;
(d) from -400 to 370, step 70 km s$^{-1}$;
North is up, and the east to the left.
}
\label{kvfmapb}
\end{figure*}

\clearpage
\begin{figure*}
\vspace{12.0 cm}
\begin{tabular}{c}
\includegraphics{fig21z.ps}\cr
\end{tabular}
\vspace{6.0 cm}
\caption {
Profile of [N\,{\sc ii}]$\lambda$6583 velocity, through the main ejection
at PA = 100$^{\circ}$ (obtained from the velocity field of Fig. 20a).
}
\label{kvperfilv}
\end{figure*}

\clearpage
\begin{figure*}
\vspace{12.0 cm}
\begin{tabular}{c}
\includegraphics{fig22az.ps} \cr
\includegraphics{fig22bz.ps} \cr
\end{tabular}
\vspace{5.0 cm}
\caption {
Smooth [N {\sc ii}]$\lambda$6583 velocity field (a) and the residuals of
the subtraction of the smooth model from the observed
[N {\sc ii}]$\lambda$6583 velocity field (b).
North is up, and the east to the left.
The scales and ranges of velocity are given in units of [km $\times$ s$^{-1}$].
}
\label{smooth}
\end{figure*}

\clearpage
\begin{figure*}
\vspace{12.0 cm}
\begin{tabular}{c}
\includegraphics{fig23az.ps} \cr
\includegraphics{fig23bz.ps} \cr
\end{tabular}
\vspace{6.0 cm}
\caption {
WHT + Integral  emission line velocity fields of the ionized gas, for:
(a) [N\,{\sc ii}]$\lambda$6583 at position 2 (i.e. centred in the nuclei);
(b) [N\,{\sc ii}]$\lambda$6583 at position 3 (i.e. centred in the H {\sc ii} region
complex).
The dashed lines show negative values.
The centre of the fibre bundle define the zero (0,0) position.
The ranges of the contours are:
(a) from -640 to 480, step 70 km s$^{-1}$;
(b) from -375 to 605, step 70 km s$^{-1}$.
North is up, and the east to the left.
}
\label{kvfmapnc}
\end{figure*}


\clearpage
\begin{figure*}
\vspace{12.0 cm}
\begin{tabular}{c}
\includegraphics{fig24az.ps} \cr
\includegraphics{fig24bz.ps} \cr
\end{tabular}
\vspace{6.0 cm}
\caption {
Emission-line ratios diagnostic diagram, of NGC\,5514, for the data
of 2D spectra of the knots in the bubble, the nuclei, and the east
complex of H {\sc ii}  regions (from Table 5).
The open symbols show the emission line ratios of 1D long--slit
spectra (from Table 3).
The regions of LINERs were adapted from the BPT diagrams published by
 Filippenko (1996) and Filippenko \&  Terlevich (1992).
}
\label{elrdiag}
\end{figure*}

\clearpage
\begin{figure*}
\vspace{12.0 cm}
\begin{tabular}{c}
\includegraphics{fig24cz.ps} \cr
\end{tabular}
\vspace{6.0 cm}
\addtocounter{figure}{-1}
\caption {Continue
}
\label{elrdiag2}
\end{figure*}

\clearpage
\begin{figure*}
\vspace{12.0 cm}
\begin{tabular}{c}
\includegraphics{fig25az.ps} \cr
\includegraphics{fig25bz.ps} \cr
\includegraphics{fig25cz.ps} \cr
\end{tabular}
\vspace{8.0 cm}
\caption {
WHT + Integral maps of emission-line ratios and width of NGC\,5514, for
position 1 (i.e. centred in the bubble).
The scale of the emission line ratio includes a factor of 100.
The FWHM are in units of km s$^{-1}$.
The position of the centre of the fibre bundle was defined as the zero (0,0)
value.
North is up, and the east to the left.
}
\label{elrmapbu}
\end{figure*}

\clearpage
\begin{figure*}
\vspace{12.0 cm}
\begin{tabular}{c}
\includegraphics{fig26az.ps} \cr
\includegraphics{fig26bz.ps} \cr
\includegraphics{fig26cz.ps} \cr
\end{tabular}
\vspace{8.0 cm}
\caption {
WHT + Integral maps of emission-line ratios and width of NGC\,5514, for
position 2 (i.e. centred in the nuclei regions).
The scale of the emission line ratio includes a factor of 100.
The FWHM are in units of km s$^{-1}$.
The position of the centre of the fibre bundle was defined as the zero (0,0)
value.
North is up, and the east to the left.
}
\label{elrmapnu}
\end{figure*}

\clearpage
\begin{figure*}
\vspace{12.0 cm}
\begin{tabular}{c}
\includegraphics{fig27az.ps} \cr
\includegraphics{fig27bz.ps} \cr
\includegraphics{fig27cz.ps} \cr
\end{tabular}
\vspace{8.0 cm}
\caption {
WHT + Integral maps of emission-line ratios and width, for
position 3 (i.e. centred in the complex of H {\sc ii} regions).
The scale of the emission line ratio includes a factor of 100.
The FWHM are in units of km s$^{-1}$.
The position of the centre of the fibre bundle was defined as the zero (0,0)
value.
North is up, and the east to the left.
}
\label{elrmapco}

\end{figure*}

\clearpage

\begin{figure*}
\vspace{12.0 cm}
\begin{tabular}{c}
\includegraphics{fig28z.ps} \cr
\end{tabular}
\vspace{3.0 cm}
\caption {
Schematic diagram of the likely orbital plane geometry,
of the NGC 5514 merger (adapted from Murphy et al. 2001).
}
\label{mergerc}
\end{figure*}

\clearpage

\begin{figure*}
\vspace{12.0 cm}
\begin{tabular}{c}
\includegraphics{fig29z.ps} \cr
\end{tabular}
\vspace{3.0 cm}
\caption {
Schematic diagram for the outflow geometry in the galactic bubble of
NGC\,5514 (adapted from Richichi \& Paresce 2003).
The line of sight to the system is approximately in the foreground of the
drawing. The axis of the bubble forms an angle of approximately 45$^{\circ}$
(a mean value) with the line of sight. The west (right) side of the bubble
is probably the nearest part, since we detected in this area a
blueshifted or approaching OF/ejection (see the text).
}
\label{bubblec}
\end{figure*}

\clearpage

\begin{figure*}
\vspace{12.0 cm}
\begin{tabular}{c}
\includegraphics{fig30z.eps} \cr
\end{tabular}
\vspace{3.0 cm}
\caption {
Plot of the ratio Fe {\sc ii}$\lambda$4570/H$\beta$ vs. Offset of the H$\beta$
broad and narrow emission line components
(probably associated with outflow; see the text).
The data were obtained from  Table 8, and for IR QSOs.
}
\label{feof}
\end{figure*}

\end{document}